\documentclass[twocolumn,english,prl]{revtex4}
\usepackage[T1]{fontenc}
\usepackage[latin9]{inputenc}
\usepackage{array}
\usepackage{units}
\usepackage{multirow}
\usepackage{amsmath}
\usepackage{amssymb}
\usepackage{graphicx}
\usepackage{esint}

\makeatletter

\providecommand{\tabularnewline}{\\}

\@ifundefined{textcolor}{}
{%
 \definecolor{BLACK}{gray}{0}
 \definecolor{WHITE}{gray}{1}
 \definecolor{RED}{rgb}{1,0,0}
 \definecolor{GREEN}{rgb}{0,1,0}
 \definecolor{BLUE}{rgb}{0,0,1}
 \definecolor{CYAN}{cmyk}{1,0,0,0}
 \definecolor{MAGENTA}{cmyk}{0,1,0,0}
 \definecolor{YELLOW}{cmyk}{0,0,1,0}
}

\usepackage{babel}

\makeatother

\usepackage{babel}
\begin{document}

\title{Mobility enhancement and temperature dependence in top-gated single-layer
MoS$_{2}$ }

\author{Zhun-Yong Ong}

\affiliation{Department of Materials Science and Engineering, University of Texas
at Dallas, 800 W Campbell Rd RL10, Richardson, TX 75080}

\author{Massimo V. Fischetti}

\email{max.fischetti@utdallas.edu}

\affiliation{Department of Materials Science and Engineering, University of Texas
at Dallas, 800 W Campbell Rd RL10, Richardson, TX 75080}
\begin{abstract}
The deposition of a high-$\kappa$ oxide overlayer is known to significantly
enhance the room-temperature electron mobility in single-layer MoS$_{2}$
(SLM) but not in single-layer graphene (SLG). We give a quantitative
account of how this mobility enhancement is due to the non-degeneracy
of the two-dimensional electron gas system in SLM at accessible temperatures.
Using our charged impurity scattering model {[}Ong and Fischetti,
\emph{Phys. Rev. B} \textbf{86}, 121409 (2012){]} and temperature-dependent
polarizability, we calculate the charged impurity-limited mobility
($\mu_{\textrm{imp}}$) in SLM with and without a high-$\kappa$ (HfO$_{2}$)
top gate oxide at different electron densities and temperatures. We
find that the mobility enhancement is larger at low electron densities
and high temperatures because of finite-temperature screening, thus
explaining the enhancement of the mobility observed at room temperature.
$\mu_{\textrm{imp}}$ is shown to decrease significantly with increasing
temperature, suggesting that the strong temperature dependence of
measured mobilities should not be interpreted as being solely due
to inelastic scattering with phonons. We also reproduce the recently
seen experimental trend in which the temperature scaling exponent
($\gamma$) of $\mu_{\textrm{imp}}\propto T^{-\gamma}$ is smaller
in top-gated SLM than in bare SLM. Finally, we show that a $\sim37$
percent mobility enhancement can be achieved by reducing the HfO$_{2}$
thickness from 20 to 2 nm.
\end{abstract}
\maketitle

\section{Introduction}

In recent years, two-dimensional metal dichalcogenides have attracted
much attention as viable alternatives to graphene\ \cite{KNovoselov:Science04}
for post-CMOS nanoelectronic applications\ \cite{QHWang:NatureNano12}.
In particular, single-layer MoS$_{2}$ (SLM) has been the focus of
much research\ \cite{BRadisavljevic:Nature11,YYoon:NL11,KKaasbjerg:PRB12,MWLin:JPD12,JLin:APL13_MoS2Overlayer,BRadisavljevic:NatMat13}.
Like single-layer graphene (SLG), SLM is an atomically thin two-dimensional
crystal. Given its atomic thickness and close proximity to the substrate,
the electron density in SLM can be tuned via a vertical electric field.
However, this means SLM is highly susceptible to the local electrical
field generated by charged impurities near or at the substrate surface.
Therefore, the electron mobility is expected to be strongly affected
by charged impurity (CI) scattering\ \cite{SAdam:SSC09} and/or remote
phonon scattering\ \cite{MVFischetti:JAP01,SFratini:PRB08,ZYOng:PRB12_IPP,ZYOng:PRB12_Erratum,ZYOng:APL13_TopOxide}.

Radisavljevic and co-workers\ \cite{BRadisavljevic:Nature11} recently
measured the electron mobility ($\mu_{e}$) in SiO$_{2}$-supported
SLM to be between 0.1 and 10 cm$^{2}$V$^{-1}$s$^{-1}$. However,
when a thin layer of HfO$_{2}$ ($\kappa=22$) was deposited on the
SLM to form a top gate, they reported a 20-fold mobility increase
of $\sim200$ cm$^{2}$V$^{-1}$s$^{-1}$ at room temperature. More
recent and accurate mobility measurements\ \cite{BRadisavljevic:NatMat13}
based on the Hall effect yield a maximum mobility of $\sim63$ cm$^{2}$V$^{-1}$s$^{-1}$
in top-gated SLM and $\sim17$ cm$^{2}$V$^{-1}$s$^{-1}$ in bare
uncovered SLM at 260 K, an almost 4-fold improvement. This mobility
enhancement was attributed to screening from the HfO$_{2}$ which
reduces CI scattering, believed to be the dominant scattering process.
Amani and co-workers also found a similar 3-fold enhancement in Al$_{2}$O$_{3}$-covered
SLM grown with chemical vapor deposition\ \cite{MAmani:APL13}.

This mobility enhancement from dielectric screening is puzzling given
that the same effect has not been seen in top-gated SLG. When Fallahazad
and co-workers deposited HfO$_{2}$ on SiO$_{2}$-supported SLG, they
did not observe any mobility enhancement although they did find that
a thinner gate oxide increases the mobility in SLG\ \cite{BFallahazad:APL10}.
This has been explained as consequence of greater screening of the
charged impurities by the metal gate\ \cite{ZYOng:PRB12_TopGate}.
In every instance that we know of\ \cite{MCLemme:SSE08,SKim:APL09,BFallahazad:APL10,JPezoldt:PSS10,KZou:PRL10},
the deposition of an oxide layer on high-mobility, \emph{non-epitaxial}
SLG has lead to a mobility \emph{decrease,} probably as a result of
more CI and defect scattering. Thus, it is surprising to observe a
several-fold improvement for SLM. This suggests that CI scattering
is qualitatively different in top-gated SLM\ \cite{BRadisavljevic:Nature11}.
The variance between the data from Refs.\ \cite{BRadisavljevic:Nature11}
and \cite{BFallahazad:APL10} is striking, and may be due to the different
electronic band structures, the nature of the interaction between
the substrate and the SLG/SLM, or the type of charge screening. In
both cases, the substrate material is SiO$_{2}$ while the gate oxide
is HfO$_{2}$ (30 nm thick in Ref.\ \cite{BRadisavljevic:Nature11}
and 11 nm in Ref.\ \cite{BFallahazad:APL10}), and the mobility measurement
methods (two-probe) are similar. This and the similar stack structure
rule out the possibility of the difference being due to the top gate
capacitance\ \cite{JLXia:NL10,MFuhrer:NatureNanotech13}.

Another salient feature of electron transport in SLM is that the deposition
of the top gate oxide alters the temperature dependence of the electron
mobility. At room temperature (300 K), the phonon-limited electron
mobility is predicted to scale as $\mu_{e}\propto T^{-\gamma}$ with
$\gamma=1.69$ and $\mu_{e}\approx410$ cm$^{2}$V$^{-1}$s$^{-1}$
in bare SLM and $\gamma=1.52$ and $\mu_{e}\approx480$ cm$^{2}$V$^{-1}$s$^{-1}$
in top-gated SLM where the homopolar optical phonon mode is assumed
to be quenched\ \cite{KKaasbjerg:PRB12}. Measurements by Radisavljevic
and Kis of the high-temperature ($T=80$ to 280 K) Hall mobility in
bare SLM yield $\gamma\approx1.4$, in good agreement with Ref.\ \cite{KKaasbjerg:PRB12},
although the absolute value of the mobility is about one order of
magnitude smaller with $\mu_{e}<20$ cm$^{2}$V$^{-1}$s$^{-1}$ at
260 K\ \cite{BRadisavljevic:NatMat13}. Their measurements on top-gated
SLM also yield $\gamma=0.3$ to 0.73 with $\mu_{e}=57$ to 63.7 cm$^{2}$V$^{-1}$s$^{-1}$
at 260 K in samples exhibiting the metal-insulator transition. Their
bare SLM results are also in good agreement with the more recent data
from Baugher and co-workers whose measurements on bare SLM give $\mu_{e}<20$
cm$^{2}$V$^{-1}$s$^{-1}$ and $\gamma=1.7$ at 300 K\ \cite{BBaugher:NL13}.
Although experimentally determined values of $\gamma$ from Refs.\ \cite{BBaugher:NL13,BRadisavljevic:NatMat13}
($\gamma=1.4$ and 1.7 respectively) agree with the theoretically
predicted value of $\gamma=1.69$ in bare SLM, the experimental values
($\mu_{e}<20$ cm$^{2}$V$^{-1}$s$^{-1}$) are one order of magnitude
smaller than the theoretical value ($\mu_{e}\approx410$ cm$^{2}$V$^{-1}$s$^{-1}$)
and suggest that intrinsic phonon scattering is not the dominant factor
in the temperature dependence of $\mu_{e}$.

In this article, we study temperature-dependent, charged impurity-limited
electron transport in bare and top-gated single-layer MoS$_{2}$ by
adapting the model developed in Ref.\ \cite{ZYOng:PRB12_TopGate}
and including not only the effect of the dielectric environment but
also the temperature dependence of the charge polarizability. HfO$_{2}$-covered
SLM on a SiO$_{2}$ substrate is used as a model system here although
the theory can be easily generalized to other gate dielectrics and
single-layer transition metal dichalcogenides (TMDs). Our use of the
temperature-dependent charge polarizability is motivated by the electron
transport data from Ghatak and co-workers\ \cite{SGhatak:ACSNano11},
which have been interpreted to imply that charged impurities are weakly
screened at room temperature. For simplicity, electron-phonon interaction
is mostly ignored here to isolate the effects of screening by the
charge polarizability as well as the dielectric environment although
scattering with the intrinsic phonons is included when it comes to
the mobility scaling with temperature. The difference between the
charge impurity-limited electron mobility ($\mu_{\textrm{imp}}$)
in bare and top-gated SLM at different temperatures ($T$) and electron
densities ($n$) is used to explain the screening effect of the gate
oxide on room-temperature electron transport. We also show that the
lower mobility at higher temperatures can be due to temperature-dependent
screening. Lastly, we predict the scaling of $\mu_{\textrm{imp}}$
with the gate oxide thickness ($t_{\textrm{ox}}$) at room temperature.

\section{Methodology}

\begin{figure}
\includegraphics[width=2.4in]{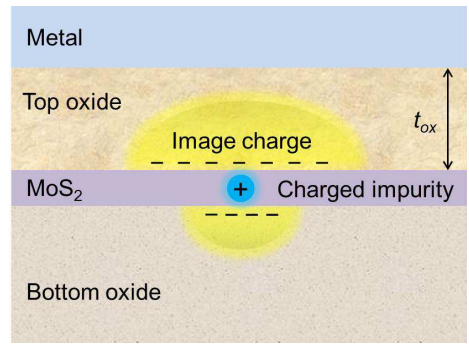}

\caption{(Color online) Basic model used in our calculation. The SLM is an
infi{}nitely thin layer at the interface ($z=0$) between a semi-infi{}nite
substrate and a top oxide layer of thickness $t_{\textrm{ox}}$. The
dielectric is capped with metal, which we assume to be a perfect conductor.
The charged impurity at the interface has image charges under and
above it in the substrate and top gate, respectively. }
\label{Fig:SetUp}
\end{figure}

\subsection{Charged impurity scattering potential}

A schematic of the setup is shown in Fig.\ \ref{Fig:SetUp}. The
model consists of a SLM sheet sandwiched between two oxide layers
with the interface at $z=0$ on the $x$-$y$ plane. The substrate
oxide (SiO$_{2}$) is semi-infinite ($z<0$) while the gate oxide
has a thickness of $t_{\textrm{ox}}$ (i.e. $0\leq z<t_{\textrm{ox}}$).
We approximate SLM as an ideal zero-thickness two-dimensional electron
gas (2DEG). To determine $\mu_{\textrm{imp}}$, we compute the scattering
rate $\Gamma_{\textrm{imp}}$ for the single CI scattering potential
$\phi_{q}^{\textrm{scr}}(0)$. The expression for the $\phi_{q}^{\textrm{scr}}(0)$
is\ \cite{ZYOng:PRB12_TopGate}: 
\begin{equation}
\phi_{q}^{\textrm{scr}}(z=0)=\frac{e^{2}G_{q}(0,0)}{\epsilon_{\textrm{2D}}(q,T)}\label{Eq:ScreenedPotential}
\end{equation}
where $\mathbf{q}$, $e$ and $G_{q}(0,0)$ are the wave vector, the
absolute electron charge quantum, and the Fourier transform (with
respect to $x$ and $y$) of the Green's function solution of the
Poisson equation, respectively; $\epsilon_{\textrm{2D}}(q,T)$ is
the generalized static dielectric function, given by $\epsilon_{\textrm{2D}}(q,T)=1-e^{2}G_{q}(0,0)\Pi(q,T,E_{F})$
where $\Pi(q,T,E_{F})$ is the temperature-dependent static charge
polarizability. The expression for $G_{q}(0,0)$ is $G_{q}(0,0)=\{[\epsilon_{\textrm{tox}}^{0}\coth(qt_{\textrm{ox}})+\epsilon_{\textrm{box}}^{0}]q\}^{-1}$
where $\epsilon_{\textrm{tox}}^{0}$ and $\epsilon_{\textrm{box}}^{0}$
are the static permittivity of the top and bottom oxides, respectively.
The electrostatic boundary conditions are included in $G_{q}(0,0)$.

\subsection{Fermi temperature and temperature-dependent screening}

\begin{figure}
\includegraphics[width=3.3in]{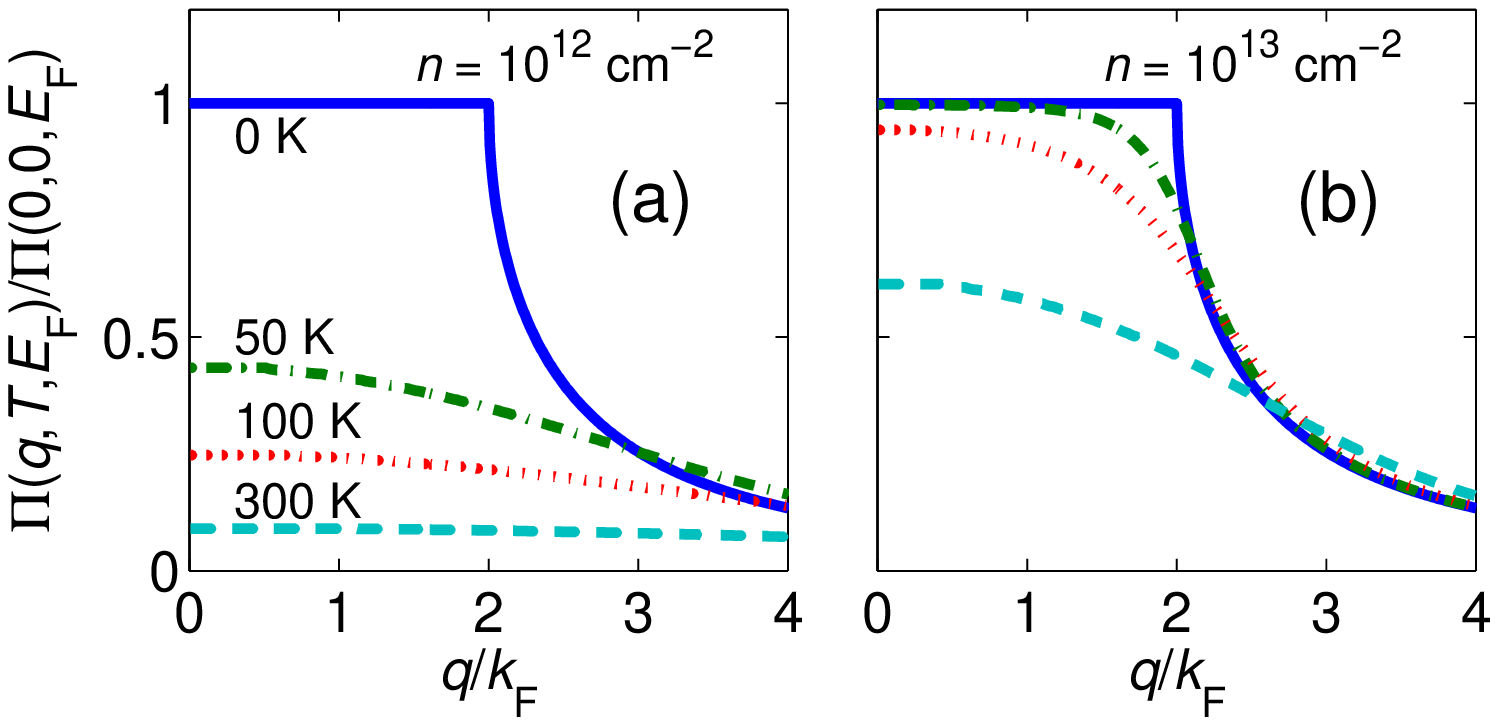}

\includegraphics[width=3.3in]{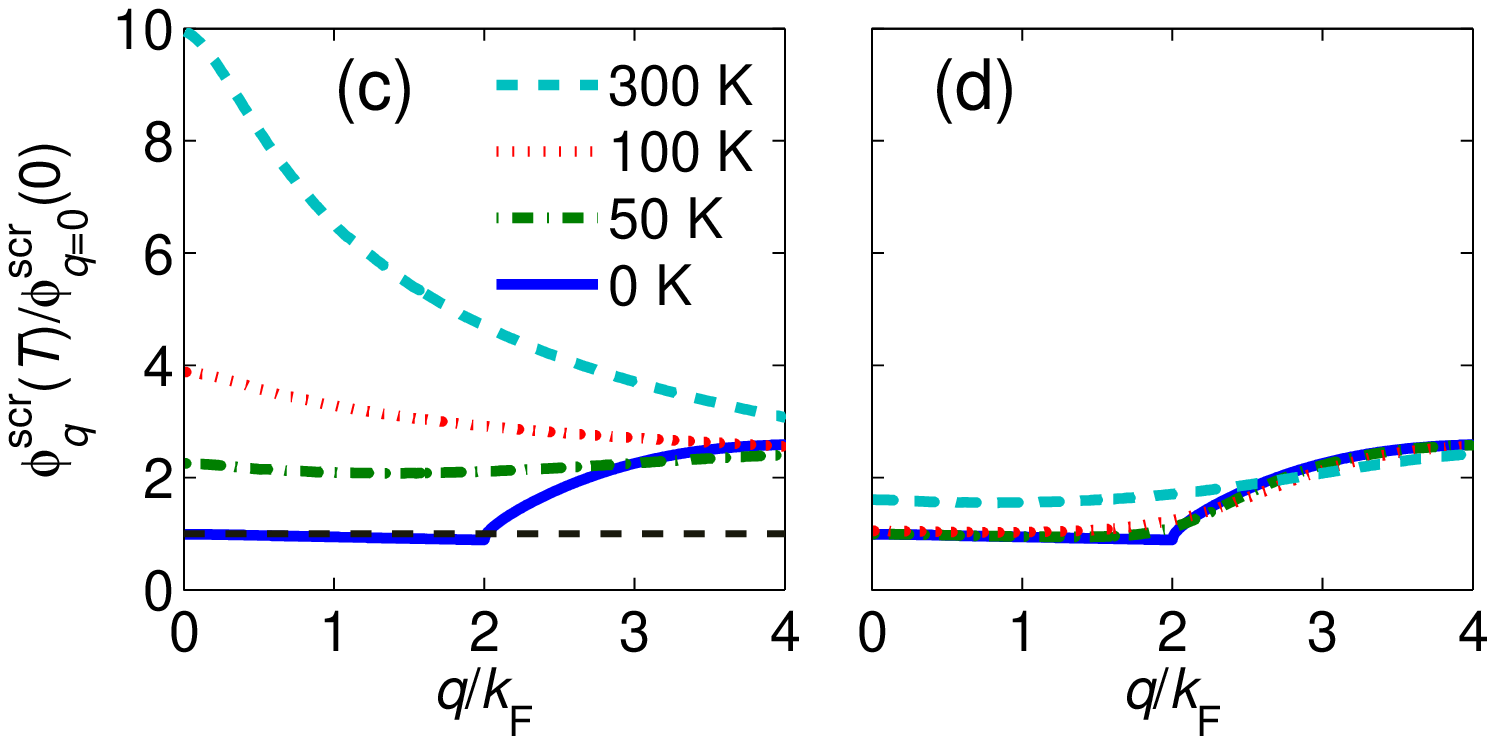}

\caption{(Color online) Plot of the normalized polarizability $\Pi(q,T,E_{F})/\Pi(0,0,E_{F})$
for (a) $n=10^{12}$ cm$^{-2}$ and (b) $n=10^{13}$ cm$^{-2}$ at
$T=0$ K (solid), 50 K (dash-dot), 100 K (dotted) and 300 K (dashed).
We also plot the corresponding normalized scattering potential $\phi_{q}^{\textrm{scr}}(T)/\phi_{q=0}^{\textrm{scr}}(T=0)$
for (c) $n=10^{12}$ cm$^{-2}$ and (d) $n=10^{13}$ cm$^{-2}$ in
top-gated SLM.}

\label{Fig:NormalizedPolarizability}
\end{figure}

While graphene remains degenerate even at low density around room
temperature, in TMDs the temperature dependence of the dielectric
response can play a significant role. We take it into account by first
examining the long-wavelength, finite-temperature approximation for
$\Pi(q,T,E_{F})$\ \cite{JHDavies:Book97}, \emph{i.e.} 
\begin{equation}
\lim_{q\rightarrow0}\Pi(q,T,E_{F})=-\frac{gm_{\textrm{eff}}}{2\pi\hbar^{2}}\left[1-\exp\bigg(\frac{-\pi\hbar^{2}n}{2m_{\textrm{eff}}k_{B}T}\bigg)\right]\ ,\label{Eq:TDependentPolarization}
\end{equation}
where $g$ and $m_{\textrm{eff}}$ are the valley-spin degeneracy
($g=4$) and the effective electron mass, respectively; $E_{F}$ is
the chemical potential and is related to $n$ via the equation $E_{F}=k_{B}T\ln\{\exp[\pi\hbar^{2}n/(2m_{\textrm{eff}}k_{B}T)]-1\}$;
$k_{B}$ and $\hbar$ are the Boltzmann and Planck constant, respectively.
For a given electron density $n$, the 2DEG can be considered degenerate
when $T\ll T_{F}$ where $T_{F}=\pi\hbar^{2}n/(2m_{\textrm{eff}}k_{B})$
is the characteristic \emph{Fermi temperature}. At $n=10^{12}$ cm$^{-2}$,
$T_{F}=29$ K. Therefore, we need to use finite-temperature screening
for the range of electron densities and temperatures in our calculations
later. At finite $q$, we can use the more general expression\ \cite{PMaldague:SurfSci1978,FStern:PRL80,TAndo:RMP82}:
\begin{equation}
\Pi(q,T,E_{F})=\int_{0}^{\infty}d\mu\frac{\Pi(q,0,\mu)}{4k_{B}T\cosh^{2}(\frac{E_{F}-\mu}{2k_{B}T})}\ ,\label{Eq:FiniteWavelengthPolarization}
\end{equation}
where $\Pi(q,0,\mu)=\Pi(0,0,\mu)\{1-\Theta(q-2k_{F})[1-(2k_{F}/q)^{2}]^{\nicefrac{1}{2}}\}$
with $k_{F}=\sqrt{2m_{\textrm{eff}}\mu}/\hbar$ and $\Pi(0,0,\mu)=-gm_{\textrm{eff}}/(2\pi\hbar^{2})$.
Figure\ \ref{Fig:NormalizedPolarizability} shows the $q$-dependence
of $\Pi(q,T,E_{F})$ at $T=0$, 50, 100 and 300 K for (a) $n=10^{12}$
cm$^{-2}$ and (b) $n=10^{13}$ cm$^{-2}$. For the same given $T$,
the change in the polarizability relative to the 0 K case is greater
at $n=10^{12}$ cm$^{-2}$ ($T_{F}=29$ K) than at $n=10^{13}$ cm$^{-2}$
($T_{F}=290$ K). We also observe that $\Pi(q,T,E_{F})$ is significantly
smaller at 300 K than at 0 K. In general, $\Pi(q,T,E_{F})$ in Eq.\ (\ref{Eq:TDependentPolarization}),
which appears in the denominator in Eq.\ (\ref{Eq:ScreenedPotential})
and corresponds to charge screening, vanishes as $n\rightarrow0$
or $T\rightarrow\infty$, \emph{i.e.} charge screening weakens with
decreasing electron density or increasing temperature. Hence, the
CI scattering strength increases as $n\rightarrow0$ or $T\rightarrow\infty$
. To illustrate this, we plot the corresponding scattering potential
$\phi_{q}^{\textrm{scr}}$ in top-gated SLM at $T=0$, 50, 100 and
300 K, normalized to $\phi_{q=0}^{\textrm{scr}}$ at $T=0$ K, in
Fig.\ \ref{Fig:NormalizedPolarizability} for (c) $n=10^{12}$ cm$^{-2}$
and (d) $n=10^{13}$ cm$^{-2}$. For $n=10^{13}$ cm$^{-2}$, the
scattering potential remains relatively unchanged as $T$ increases,
unlike the scattering potential for $n=10^{12}$ cm$^{-2}$ which
increases by up to an order of magnitude as $T$ increases from 0
K to 300 K, because the Fermi temperature at $n=10^{13}$ cm$^{-2}$
is $T_{F}=290$ K.

Following Ref.\ \cite{KKaasbjerg:PRB12} we approximate the electron
dispersion in SLM with a parabolic expression $E(\mathbf{k})=\hbar^{2}k^{2}/(2m_{\textrm{eff}})$
with effective mass $m_{\textrm{eff}}=0.48m_{0}$ (where $m_{0}$
is the free electron mass) and minimum at the symmetry point \emph{K}.
The use of a single valley should not constitute a big error since
at low fields no interband transitions are expected to take place\ \cite{KKaasbjerg:PRB12}.
The expression for the CI scattering rate is\ \cite{ZYOng:PRB12_TopGate}:
\begin{eqnarray}
\Gamma_{\textrm{imp}}(E_{\mathbf{k}}) & = & \frac{n_{\textrm{imp}}}{2\pi\hbar}\int d\mathbf{k}'|\phi_{|\mathbf{k-k'}|}^{\textrm{scr}}(d)|^{2}\nonumber \\
 &  & \times(1-\cos\theta_{\mathbf{kk'}})\delta(E_{\mathbf{k}}-E_{\mathbf{k}'})\ ,\label{Eq:ImpurityScatteringRate}
\end{eqnarray}
where $\theta_{\mathbf{kk'}}$ is the scattering angle between the
$\mathbf{k}$ and $\mathbf{k}'$ states, and $n_{\textrm{imp}}$ is
the CI concentration which is a fitting parameter. The expression
for the CI-limited electron mobility is: 
\begin{equation}
\mu_{\textrm{imp}}=\frac{e}{\pi\hbar^{2}k_{B}T}\int_{0}^{\infty}f(E)[1-f(E)]\Gamma_{\textrm{imp}}(E)^{-1}E\ dE\ ,\label{Eq:ImpurityLimitedMobility}
\end{equation}
where $f(E)$ is the equilibrium Fermi-Dirac distribution function.
By using Eq.\ (\ref{Eq:ImpurityLimitedMobility}), we assume that
electron transport is described by semiclassical band transport, as
in Refs.\ \cite{KKaasbjerg:PRB12,XLi:PRB13_IntrinsicMoS2} and opposed
to hopping transport as suggested in Ref.\ \cite{SGhatak:ACSNano11},
and that the dominant scattering mechanism is CI scattering, which
is mostly at the Fermi surface. The main momentum relaxation process
corresponds to the momentum change of $q\sim2k_{F}$ and the related
Fourier component of the scattering potential $\phi_{2k_{F}}^{\textrm{scr}}$,
which is inversely proportional to the dielectric function $\epsilon_{\textrm{2D}}(2k_{F},T)$
and strongly affected by the temperature broadening of $\Pi(2k_{F},T,E_{F})$\ \cite{FStern:PRL80,SDasSarma:PRB85}.

\section{Results and Discussion}

\subsection{Electron density dependence of mobility at low and high temperature}

\begin{figure}
\includegraphics[width=3.3in]{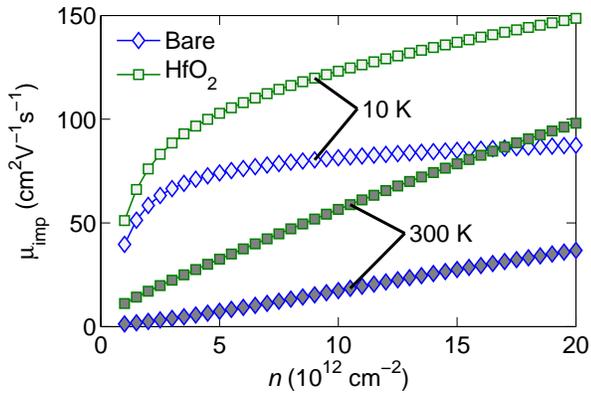}

\caption{(Color online) Plot of $\mu_{\textrm{imp}}^{0}$ (`Bare') and $\mu_{\textrm{imp}}^{\textrm{TG}}$
(`HfO$_{2}$') at $T=10$ K (hollow symbols) and $T=300$ K (solid
symbols) for $n=10^{12}$ to $2\times10^{13}$ cm$^{-2}$ for $n_{\textrm{imp}}=4.0\times10^{12}$
cm$^{-2}$. At 300 K, the mobility scales almost linearly with the
electron density.}
\label{Fig:ScreenedMobility}
\end{figure}

The variables $\mu_{\textrm{imp}}^{0}$ and $\mu_{\textrm{imp}}^{\textrm{TG}}$
denote the CI-limited mobility in bare SiO$_{2}$-supported ($\epsilon_{\textrm{tox}}^{0}=\epsilon_{0}$
and $t_{\textrm{ox}}=\infty$) and 30-nm-HfO$_{2}$-top-gated, SiO$_{2}$-supported
($\epsilon_{\textrm{tox}}^{0}=22\epsilon_{0}$ and $t_{\textrm{ox}}=30$
nm) SLM. We assume $n_{\textrm{imp}}=4\times10^{12}$ cm$^{-2}$ for
ease of comparison with the results in Ref.\ \cite{BRadisavljevic:Nature11}.
At $n=2\times10^{13}$ cm$^{-2}$ and $T=10$ K, this yields $\mu_{\textrm{imp}}^{\textrm{TG}}\sim150$
cm$^{2}$V$^{-1}$s$^{-1}$, comparable to that measured by Radisavljevic
and Kis\ \cite{BRadisavljevic:NatMat13} at low temperatures. We
first calculate and plot in Fig.\ \ref{Fig:ScreenedMobility} $\mu_{\textrm{imp}}^{0}$
and $\mu_{\textrm{imp}}^{\textrm{TG}}$ at $T=10$ K, from $n=10^{12}$
to $2\times10^{13}$ cm$^{-2}$ in steps of $\Delta n=10^{12}$ cm$^{-2}$.
The corresponding Fermi temperature range is $T_{F}=29$ to 580 K.
We find that both $\mu_{\textrm{imp}}^{0}$ and $\mu_{\textrm{imp}}^{\textrm{TG}}$
increase monotonically with $n$, in good agreement with the Hall
mobility data given in Ref.\ \cite{BBaugher:NL13}, with the density
dependence stronger for $\mu_{\textrm{imp}}^{\textrm{TG}}$. At low
densities ($n<4\times10^{12}$ cm$^{-2}$), the density dependence
is markedly greater. 

Our results indicate that $\mu_{\textrm{imp}}^{\textrm{TG}}$ is higher
than $\mu_{\textrm{imp}}^{0}$, with the relative difference increasing
with $n$; at $n=10^{12}$ cm$^{-2}$, we have $\mu_{\textrm{imp}}^{\textrm{TG}}/\mu_{\textrm{imp}}^{0}=1.29$
while at $n=2\times10^{13}$ cm$^{-2}$, we have $\mu_{\textrm{imp}}^{\textrm{TG}}/\mu_{\textrm{imp}}^{0}=1.70$.
This suggests that the mobility enhancement from overlaying SLM with
a high-$\kappa$ material is modest at low temperatures. This is because
at low temperatures ($T\ll T_{F}$), screening is dominated by the
charge polarizability. To see how, we rewrite the scattering potential
of a single CI in Eq.\ (\ref{Eq:ScreenedPotential}) as: 
\[
\phi_{q}^{\textrm{scr}}(0)=\frac{e^{2}G_{q}(0,0)}{1-e^{2}G_{q}(0,0)\Pi(q,T,E_{F})}\ .
\]
In the long-wavelength limit, the second term in the denominator,
which corresponds to the screening charge, dominates, giving us $\lim_{q\rightarrow0}\phi_{q}^{\textrm{scr}}(0)=-\Pi(q=0,T,E_{F})^{-1}$.
Thus, the scattering potential is independent of the dielectric environment
in the long-wavelength limit and depends only on the polarizability.
At low $T$ and $q<2k_{F}$, the polarizability is nearly independent
of $q$, \emph{i.e.} $\lim_{T\rightarrow0}\Pi(q<2k_{F},T,E_{F})=-2m_{\textrm{eff}}/(\pi\hbar^{2})$.
This explains why $\mu_{\textrm{imp}}^{\textrm{TG}}/\mu_{\textrm{imp}}^{0}$
is close to unity. The decrease of $\mu_{\textrm{imp}}^{0}$ and $\mu_{\textrm{imp}}^{\textrm{TG}}$
with smaller $n$ is due to the fact that at small $n$, we have $\lim_{n\rightarrow0}\Pi(q,T,E_{F})\propto n$
which implies that the scattering potential strength scales as $\sim n^{-1}$.

\begin{figure}
\includegraphics[width=3.3in]{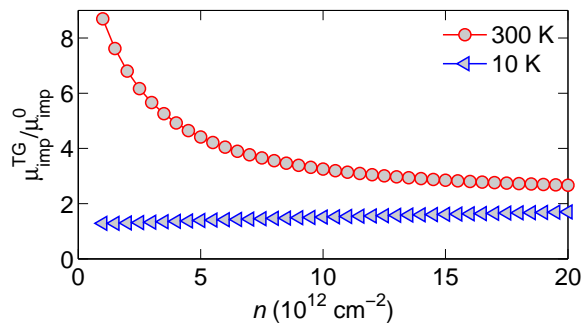}

\caption{(Color online) Plot of the mobility enhancement $\mu_{\textrm{imp}}^{\textrm{TG}}/\mu_{\textrm{imp}}^{0}$
at 10 K (circle) and 300 K (triangle). The mobility enhancement increases
at higher temperatures or lower electron densities. }
\label{Fig:MobilityEnhancement}
\end{figure}

On the other hand, experimental measurements reveal that covering
SLM with a high-$\kappa$ dielectric leads to significant \emph{room-temperature}
mobility enhancement\ \cite{BRadisavljevic:NatMat13}. This suggests
that screening by the dielectric plays a greater role in the temperature
regime $T\gtrsim T_{F}$. Hence, the screening effect of the charge
polarizability in SLM is less significant. To show this, we repeat
our calculation of $\mu_{\textrm{imp}}^{0}$ and $\mu_{\textrm{imp}}^{\textrm{TG}}$
but now at room temperature (300 K). The room-temperature results
are also shown in Fig.\ \ref{Fig:ScreenedMobility}. In contrast
to the low-temperature results in Fig.\ \ref{Fig:ScreenedMobility},
$\mu_{\textrm{imp}}^{0}$ and $\mu_{\textrm{imp}}^{\textrm{TG}}$
are strongly density-dependent and scale almost linearly with $n$,
in good agreement with the room-temperature data for bare SLM by Ghatak
and co-workers\ \cite{SGhatak:ACSNano11}. The linear density-dependence
is a signature of weak or absent screening by the polarization charge
in SLM. Thus, the role of screening by the surrounding dielectric
media becomes more important. At low $n$, $\mu_{\textrm{imp}}^{\textrm{TG}}$
is significantly larger than $\mu_{\textrm{imp}}^{0}$. At $n=10^{12}$
cm$^{-2}$, $\mu_{\textrm{imp}}^{0}\approx1.3$ cm$^{2}$V$^{-1}$s$^{-1}$
while $\mu_{\textrm{imp}}^{\textrm{TG}}\approx11.2$ cm$^{2}$V$^{-1}$s$^{-1}$,
nearly an order-of-magnitude increase. This agrees very well with
the measured several-fold mobility enhancement reported in Refs.\ \cite{MAmani:APL13,BRadisavljevic:NatMat13}.
We plot the mobility enhancement $\mu_{\textrm{imp}}^{\textrm{TG}}/\mu_{\textrm{imp}}^{0}$
in Fig.\ \ref{Fig:MobilityEnhancement} at 10 and 300 K. The mobility
enhancement is much greater at 300 K than at 10 K because of the temperature-induced
weakening of the charge polarizability. At 300 K, the mobility enhancement
decreases and converges to that at 10 K as $n$ increases because
charge screening becomes stronger at higher densities.

\subsection{Temperature dependence of electron mobility}

The temperature dependence of the electron mobility in experiments
is often used to determine the nature of electron transport in semiconductors.
When the mobility decreases with increasing $T$, it is commonly interpreted
to be a signature of phonon-limited electron transport in the metallic
phase\ \cite{KKaasbjerg:PRB12,BBaugher:NL13,DJariwala:APL13,BRadisavljevic:NatMat13};
in the insulating phase, the rise in mobility with increasing $T$
is usually characterized as originating from hopping transport\ \cite{SGhatak:ACSNano11}.
Kaasbjerg and co-workers predict the \emph{intrinsic} phonon-limited
mobility to vary as $\mu_{e}\propto T^{-\gamma}$ ($\gamma=1.52$)
in top-gated SLM. Measurements of $\gamma$ by Radisavljevic and Kis
have it varying between 0.3 and 0.73\ \cite{BRadisavljevic:NatMat13},
which is suggestive of phonon-limited transport. For ease of comparison,
we summarize the representative theoretical and experimental mobility
results from Refs.\ \cite{KKaasbjerg:PRB12,XLi:PRB13_IntrinsicMoS2,BBaugher:NL13,BRadisavljevic:NatMat13}
in Table\ \ref{Table:MobilityComparison}, together with our results.
However, Li and co-workers\ \cite{XLi:PRB13_IntrinsicMoS2} and Kaasbjerg
and co-workers\ \cite{KKaasbjerg:PRB12} predict the \emph{K} valley-dominated,
intrinsic phonon-limited mobility values to be around several hundred
cm$^{2}$V$^{-1}$s$^{-1}$ at room temperature, which are at least
an order-of-magnitude larger than measurements\ \cite{BBaugher:NL13,BRadisavljevic:NatMat13}.
Thus, the temperature dependence of the measured mobility is probably
due to extrinsic factors such as charged impurities and remote phonons. 

The disparity between our calculated low- and room-temperature $\mu_{\textrm{imp}}$
implies that CI scattering is strongly temperature-dependent and plays
an important role in the overall mobility temperature dependence.
Hence, it is important to quantify the temperature dependence in our
model for direct comparison with experiments, in order to understand
the causes of this temperature dependence. In particular, we are interested
in the temperature scaling of the high-temperature electron mobility
($\mu_{e}\propto T^{-\gamma}$), which has been investigated theoretically
and experimentally in Refs.\ \cite{KKaasbjerg:PRB12,XLi:PRB13_IntrinsicMoS2,BBaugher:NL13,BRadisavljevic:NatMat13},
and the difference in this temperature scaling between bare and top-gated
SLM. Radisavljevic and Kis recently reported a substantial decrease
in $\gamma$, from $\gamma=1.47$ in bare SLM to $\gamma=0.3-0.73$
in top-gated SLM\ \cite{BRadisavljevic:NatMat13}, much greater than
that expected from the quenching of homopolar optical phonons\ \cite{KKaasbjerg:PRB12}.
By studying the difference in the temperature dependence of the mobility
in bare and top-gated SLM with our model, we hope to shed light on
this phenomenon. 

Since the temperature variation of the electron mobility may depend
on scattering with phonons, we compute the CI/phonon-limited electron
mobility $\mu_{e}=\mu_{\textrm{imp+phon}}$, taking into account charged
impurity as well as intrinsic phonon scattering, in addition to the
computation of the CI-limited mobility $\mu_{\textrm{imp}}$. The
intrinsic electron-phonon interactions include the longitudinal acoustic
(LA), the transverse acoustic (TA), the intervalley longitudinal optical
(LO) and the intravalley homopolar optical (HP) phonons, with the
scattering rate formulas and parameters taken directly from Ref.\ \cite{KKaasbjerg:PRB12}.
In our calculation of the CI/phonon-limited electron mobility ($\mu_{\textrm{imp+phon}}^{0}$)
in bare SLM, we include electron scattering with the LA, TA, LO and
HP phonons while in top-gated SLM, we assume that the HP phonons are
quenched (as in Ref.\ \cite{KKaasbjerg:PRB12}) and we do not include
them in our calculation of the mobility ($\mu_{\textrm{imp+phon}}^{\textrm{TG}}$). 

\begin{figure}
\includegraphics[width=3.3in]{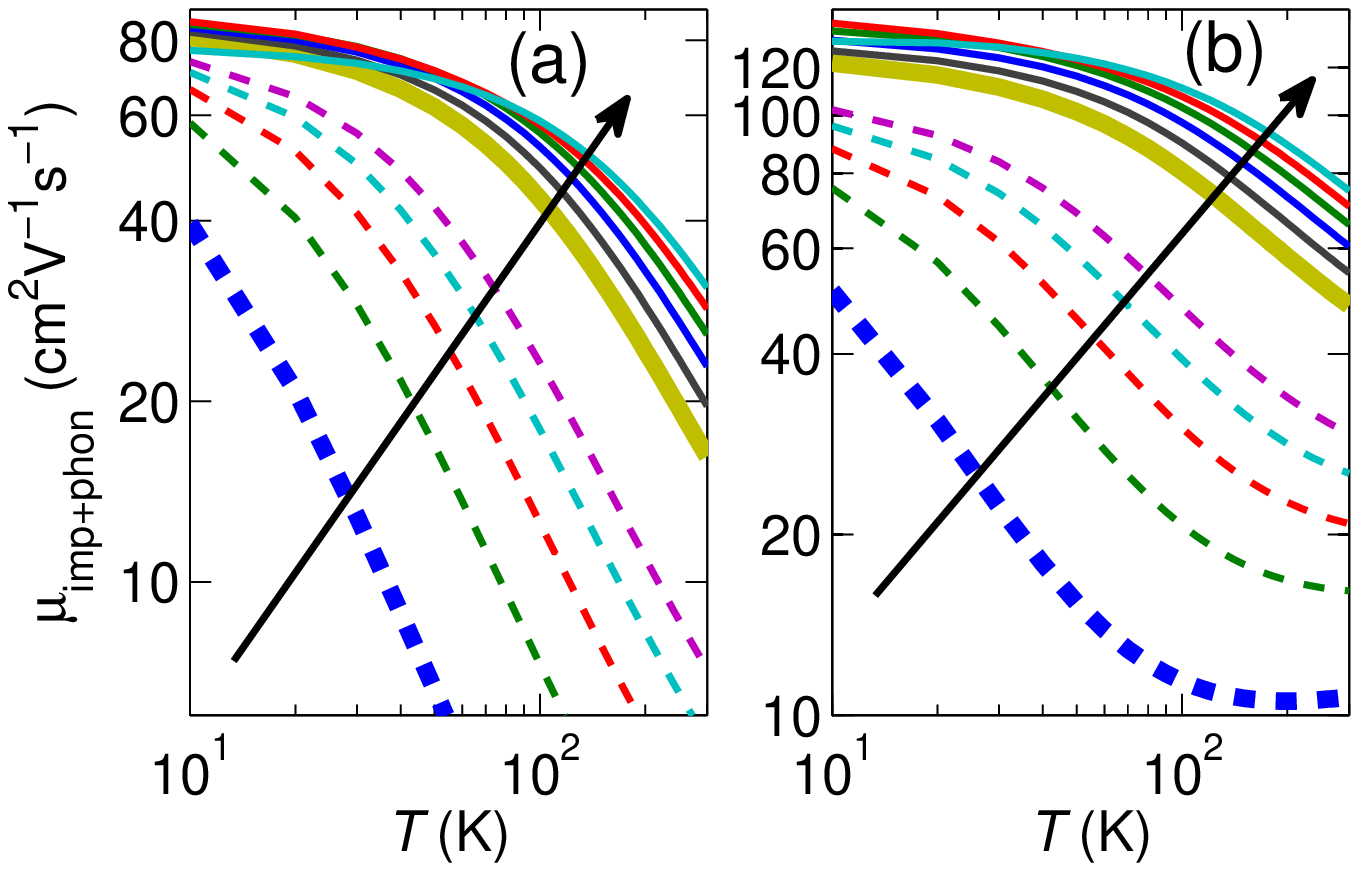}

\includegraphics[width=3.3in]{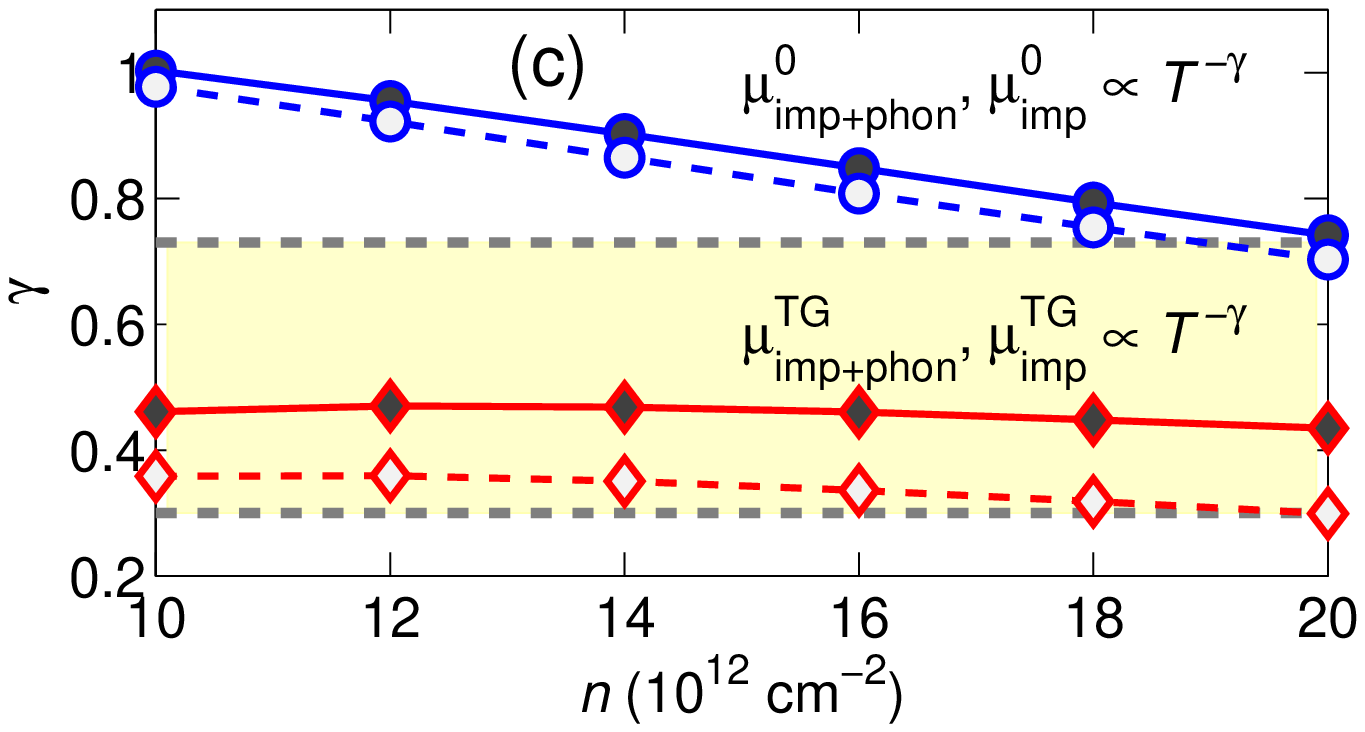}

\caption{(Color online) Plot of (a) $\mu_{\textrm{imp+phon}}^{0}$ and (b)
$\mu_{\textrm{imp+phon}}^{\textrm{TG}}$ for $n=10^{12}$ to $5\times10^{12}$
cm$^{-2}$ in steps of $\Delta n=10^{12}$ cm$^{-2}$(dashed lines)
and $n=10^{13}$ to $2\times10^{13}$ cm$^{-2}$ in steps of $\Delta n=2\times10^{12}$
cm$^{-2}$(solid lines) from $T=10$ to 300 K. The arrows indicate
the direction of increasing $n$. The thick dashed (solid) line corresponds
to $n=10^{12}$ cm$^{-2}$ ($10^{13}$ cm$^{-2}$). As $n$ increases,
$|d\mu_{\textrm{imp+phon}}/dT$| becomes smaller. (c) Plot of the
exponent $\gamma$ for $\mu_{\textrm{imp+phon}}^{0}$ (solid circles),
$\mu_{\textrm{imp}}^{0}$ (open circles), $\mu_{\textrm{imp+phon}}^{\textrm{TG}}$
(solid diamonds) and $\mu_{\textrm{imp}}^{\textrm{TG}}$ (open diamonds)
from fitting to $\mu_{e}\propto T^{-\gamma}$ over the range $T=200$
to 300 K. The shaded region bounded by the dashed lines covers the
range of $\gamma$ values (0.3 to 0.73) extracted for top-gated SLM
in Ref.\ \cite{BRadisavljevic:NatMat13}. }
\label{Fig:TemperatureScaling}
\end{figure}

Figure\ \ref{Fig:TemperatureScaling} shows (a) $\mu_{\textrm{imp+phon}}^{0}$
and (b) $\mu_{\textrm{imp+phon}}^{\textrm{TG}}$ for $n=10^{12}$
to $5\times10^{12}$ cm$^{-2}$ in steps of $\Delta n=10^{12}$ cm$^{-2}$,
which we take to be representative of the low-density regime, and
$n=10^{13}$ to $2\times10^{13}$ cm$^{-2}$ in steps of $\Delta n=2\times10^{12}$
cm$^{-2}$, which we take to be representative of the high-density
`metallic' regime, from $T=10$ to 300 K. Figures\ \ref{Fig:TemperatureScaling}(a)
and (b) show that the relative variation of the mobility with $T$
increases as $n$ becomes smaller. The decrease in $\mu_{\textrm{imp+phon}}^{0}$
is very large as we go from 10 to 300 K. For example, at $n=10^{12}$
cm$^{-2}$, $\mu_{\textrm{imp+phon}}^{0}$ decreases by $>97$ percent.
The sensitivity to changes in temperature is significantly greater
for $\mu_{\textrm{imp+phon}}^{0}$ than $\mu_{\textrm{imp+phon}}^{\textrm{TG}}$.
The corresponding results for the CI-limited mobilities ($\mu_{\textrm{imp}}^{0}$
and $\mu_{\textrm{imp}}^{\textrm{TG}}$ ) are not shown here since
they exhibit a similar trend with respect to temperature change.

From $T=200$ to $300$ K and $n=10^{13}$ to $2\times10^{13}$ cm$^{-2}$,
$\mu_{\textrm{imp+phon}}^{0}$ and $\mu_{\textrm{imp+phon}}^{\textrm{TG}}$
exhibit a power-law dependence on $T$, \emph{i.e.} $\mu_{\textrm{imp+phon}}\propto T^{-\gamma}$,
similar to that reported in Refs.\ \cite{BBaugher:NL13,BRadisavljevic:NatMat13}.
We plot $\gamma$ as a function of $n$ for $\mu_{\textrm{imp+phon}}^{0}$,
$\mu_{\textrm{imp+phon}}^{\textrm{TG}}$, $\mu_{\textrm{imp}}^{0}$
and $\mu_{\textrm{imp}}^{\textrm{TG}}$ in Fig.\ \ref{Fig:TemperatureScaling}(c).
The exponent $\gamma$ decreases with $n$ and is also much larger
for $\mu_{\textrm{imp+phon}}^{0}$ ($\gamma=$ 0.75 to 1.0) than for
$\mu_{\textrm{imp+phon}}^{\textrm{TG}}$ ($\gamma=$ 0.43 to 0.47),
in excellent agreement with Ref.\ \cite{BRadisavljevic:NatMat13}
where a significant decrease in $\gamma$ was found for top-gated
SLM. The $\gamma$ values for $\mu_{\textrm{imp+phon}}^{0}$ are comparable
to the $T^{-1}$ behavior expected for a dilute, high-temperature
2DEG\ \cite{SDasSarma:PRB85} but lower than the $\gamma=1.7$ and
1.4 from Refs.\ \cite{BBaugher:NL13} and \cite{BRadisavljevic:NatMat13},
respectively. The values for $\mu_{\textrm{imp+phon}}^{\textrm{TG}}$
are however within the range measured for top-gated SLM samples ($\gamma=$0.3
to 0.73)\ \cite{BRadisavljevic:NatMat13}. The range of $\gamma$
values for the CI-limited mobilities $\mu_{\textrm{imp}}^{0}$ and
$\mu_{\textrm{imp}}^{\textrm{TG}}$ are slightly smaller (0.70 to
0.98 and 0.30 to 0.36 respectively in the case of $\mu_{\textrm{imp}}^{0}$
and $\mu_{\textrm{imp}}^{\textrm{TG}}$) since the temperature dependence
only comes from the finite-temperature charge polarizability. Nevertheless,
we observe a similar decrease in $\gamma$ when comparing $\mu_{\textrm{imp}}^{\textrm{TG}}$
to $\mu_{\textrm{imp}}^{0}$. This implies that the change in $\gamma$
is due to the modification of CI scattering in top-gated SLM. 

In Ref.\ \cite{BRadisavljevic:NatMat13}, $\gamma$ increases with
$n$ (from $\gamma=0.55$ at $n=0.76\times10^{13}$ cm$^{-2}$ to
$\gamma=0.78$ at $n=1.35\times10^{13}$ cm$^{-2}$) in contrast to
our results for $\mu_{\textrm{imp}}^{\textrm{TG}}$ and $\mu_{\textrm{imp+phon}}^{\textrm{TG}}$
where $\gamma$ decreases as $n$ increases. This suggests that other
more strongly temperature-dependent scattering processes may be involved.
In Fig.\ \ref{Fig:ScreenedMobility}(a) and (b), $\mu_{\textrm{imp}}^{\textrm{TG}}$
increases with $n$, \emph{i.e.} CI scattering becomes less important
at higher densities. Hence, the relative contribution of the other
scattering processes may become more significant. 

\begin{table*}
\begin{tabular}{lcccccc}
\hline 
 &  &  & \multicolumn{2}{c}{Bare} & \multicolumn{2}{c}{Top-gated}\tabularnewline
\cline{4-7} 
Reference & $T$ (K) & Method & $\mu_{e}$  & $\gamma$  & $\mu_{e}$  & $\gamma$ \tabularnewline
\hline 
\hline 
Kaasbjerg \emph{et al.} (phonon-limited)\emph{\ }\cite{KKaasbjerg:PRB12} & 300 & Theory & 410 & 1.69 & 480 & 1.52\tabularnewline
Li \emph{et al.} (phonon-limited)\ \cite{XLi:PRB13_IntrinsicMoS2} & 300 & Theory & 320 & -- & -- & --\tabularnewline
Baugher \emph{et al.}\ \cite{BBaugher:NL13} & 300 & Expt. & <20 & 1.7 & -- & --\tabularnewline
Radisavljevic and Kis\ \cite{BRadisavljevic:NatMat13} & 260 & Expt. & 17.2 & 1.4 & 56.9 to 63 & 0.3 to 0.73\tabularnewline
Ong and Fischetti (CI-limited) & \multirow{2}{*}{300} & \multirow{2}{*}{Theory} & 17.4 & 0.98 & 56.5 & 0.36\tabularnewline
Ong and Fischetti (CI/phonon-limited) &  &  & 16.2 & 1.0 & 48.9 & 0.46\tabularnewline
\hline 
\end{tabular}

\caption{Comparison of \emph{representative} electron mobility $\mu_{e}$ (in
units of cm$^{2}$V$^{-1}$s$^{-1}$) and power-law exponent $\gamma$
(where $\mu_{e}\propto T^{-\gamma}$) values for bare and HfO$_{2}$
top-gated SLM from Refs.\ \cite{KKaasbjerg:PRB12,XLi:PRB13_IntrinsicMoS2,BBaugher:NL13,BRadisavljevic:NatMat13}.
The results from Li \emph{et al.}\ \cite{XLi:PRB13_IntrinsicMoS2}
and Kaasbjerg \emph{et al.}\ \cite{KKaasbjerg:PRB12} assume \emph{K}
valley-dominated, intrinsic phonon-limited electron transport. The
CI-limited results by Ong and Fischetti are computed with an impurity
concentration of $n_{\textrm{imp}}=4\times10^{12}$ cm$^{-2}$ at
the electron density of $n=10^{13}$ cm$^{-2}$ while the CI/phonon-limited
results are computed using the same $n_{\textrm{imp}}$ and phonon
parameters from Ref.\ \cite{KKaasbjerg:PRB12}. Our CI-limited mobility
results show that a significant temperature dependence can arise even
in the absence of phonon scattering.}

\label{Table:MobilityComparison}
\end{table*}

\subsection{Gate Oxide Thickness Dependence}

\begin{figure}
\includegraphics[width=3.3in]{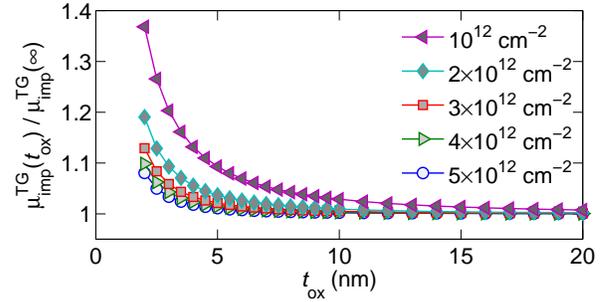}

\caption{(Color online) Plot of $\mu_{\textrm{imp}}^{\textrm{TG}}(t_{\textrm{ox}})/\mu_{\textrm{imp}}^{\textrm{TG}}(\infty)$
for $n=10^{12}$ to $5\times10^{12}$ cm$^{-2}$ and $t_{\textrm{ox}}=2$
to 20 nm at 300 K. }

\label{Fig:OxideThicknessScaling}
\end{figure}

Having shown that screening by the top gate enhances the mobility
at room temperature and low $n$ i.e. when $T\gg T_{F}$, we explore
the possibility of using a thinner gate oxide to screen the charged
impurities. We compute $\mu_{\textrm{imp}}^{\textrm{TG}}$ for $n=10^{12}$
to $5\times10^{12}$ cm$^{-2}$ and $t_{\textrm{ox}}=2$ to $20$
nm at 300 K. Figure\ \ref{Fig:OxideThicknessScaling} shows the calculated
$\mu_{\textrm{imp}}^{\textrm{TG}}$ values normalized to the $\mu_{\textrm{imp}}^{\textrm{TG}}$
for a semi-infinite top oxide layer. As expected, $\mu_{\textrm{imp}}^{\textrm{TG}}$
increases as $t_{\textrm{ox}}$ decreases because a thinner oxide
places the image charges in the metal closer to the SLM and screens
the charged impurities more effectively. At $n=10^{12}$ cm$^{-2}$,
a 37 percent enhancement in $\mu_{\textrm{imp}}^{\textrm{TG}}$ can
be achieved by reducing $t_{\textrm{ox}}$ from 20 to 2 nm. This implies
that reducing $t_{\textrm{ox}}$ can significantly mitigate the effects
of charged impurities especially when $T\gg T_{F}$.

\section{Further discussion and summary}

The underlying physics of our findings stems from the transition of
the 2DEG in SLM from degeneracy to non-degeneracy at higher \emph{accessible}
temperatures. At high temperatures ($T\gtrsim T_{F}$), charge screening
within the 2DEG becomes weaker with increasing temperature, and the
charged impurity-limited mobility becomes more dependent on screening
by the dielectric environment of the SLM. The non-degeneracy-to-degeneracy
transition also explains why the mobility enhancement is not seen
in top-gated SLG, the question posed at the beginning the paper. The
linear band structure of SLG ensures that it remains degenerate even
at room temperature. For example, the Fermi temperature in SLG exceeds
1300 K at $n=10^{12}$ cm$^{-2}$ whereas the corresponding Fermi
temperature in SLM is 29 K. Thus, charge screening within SLG is effectively
temperature-independent and dominates the screening of charged impurities
at accessible temperatures. On the other hand, charge screening within
SLM weakens with temperature and allows screening by the dielectric
environment to play a bigger role at high temperatures. 

We also point out that mobility enhancement has been observed in top-gated
epitaxial SLG\ \cite{JRobinson:ACSNano10,MHollander:NL11}. However,
it is known that the band structure of epitaxial graphene is unlike
that of ideal exfoliated SLG as a result of the formation of a substrate-induced
band gap\ \cite{SYZhou:NatMat07,SYZhou:NatMat08}. Assuming that
electron transport in epitaxial SLG is limited by CI scattering, the
low mobility in epitaxial SLG (relative to exfoliated SLG) suggests
that its intrinsic charge screening is weakened, possibly from the
aforementioned band structure modification. 

With regard to our results, we have calculated the charged impurity-limited
mobility ($\mu_{\textrm{imp}}$) in SLM with electron density and
temperature-dependent screening. Our results agree with the several-fold
improvement in room-temperature mobility reported in Refs.\ \cite{BRadisavljevic:NatMat13}
and \cite{MAmani:APL13} when a high-$\kappa$ overlayer is introduced,
and they are consistent with the weak charge screening found in Ref.\ \cite{SGhatak:ACSNano11}.
We have found that $\mu_{\textrm{imp}}$ decreases with increasing
temperature primarily as a result of temperature-dependent polarizability,
suggesting that this temperature-dependent phenomenon is not necessarily
a signature of phonon scattering. Our model also qualitatively reproduces
the change in the temperature scaling of $\mu_{e}$ when HfO$_{2}$
is deposited on SLM\ \cite{BRadisavljevic:NatMat13}. However, we
are unable to reproduce accurately the magnitude \emph{and} temperature-scaling
exponent $\gamma$ of the mobility in our model, even with the inclusion
of intrinsic phonons. This suggests that other scattering mechanisms,
possibly remote phonons\ \cite{MVFischetti:JAP01,ZYOng:PRB12_IPP,ZYOng:PRB12_Erratum,ZYOng:PRB13_RemotePhonon},
must be accounted for in a more realistic model of electron transport
in bare and top-gated SLM. Lastly, we have shown that a thinner top
oxide can lead to a significant improvement in $\mu_{\textrm{imp}}$
at low electron densities for temperatures greater than the Fermi
temperature. Our results highlight a possible strategy to optimize
the device geometry for superior electron transport properties in
single-layer MoS$_{2}$ and other transition metal dichalcogenides. 

\textit{\emph{}}

We gratefully acknowledge the support provided by Texas Instruments,
the Semiconductor Research Corporation (SRC), the South-West Academy
of Nanotechnology (SWAN) under Task 4.3 Theme 2400.011, and Samsung
Electronics Ltd. 

\bibliographystyle{apsrev4-1}
\bibliography{TopGatedMoS2Refs}

\begin{thebibliography}{37}%
\makeatletter
\providecommand \@ifxundefined [1]{%
 \@ifx{#1\undefined}
}%
\providecommand \@ifnum [1]{%
 \ifnum #1\expandafter \@firstoftwo
 \else \expandafter \@secondoftwo
 \fi
}%
\providecommand \@ifx [1]{%
 \ifx #1\expandafter \@firstoftwo
 \else \expandafter \@secondoftwo
 \fi
}%
\providecommand \natexlab [1]{#1}%
\providecommand \enquote  [1]{``#1''}%
\providecommand \bibnamefont  [1]{#1}%
\providecommand \bibfnamefont [1]{#1}%
\providecommand \citenamefont [1]{#1}%
\providecommand \href@noop [0]{\@secondoftwo}%
\providecommand \href [0]{\begingroup \@sanitize@url \@href}%
\providecommand \@href[1]{\@@startlink{#1}\@@href}%
\providecommand \@@href[1]{\endgroup#1\@@endlink}%
\providecommand \@sanitize@url [0]{\catcode `\\12\catcode `\$12\catcode
  `\&12\catcode `\#12\catcode `\^12\catcode `\_12\catcode `\%12\relax}%
\providecommand \@@startlink[1]{}%
\providecommand \@@endlink[0]{}%
\providecommand \url  [0]{\begingroup\@sanitize@url \@url }%
\providecommand \@url [1]{\endgroup\@href {#1}{\urlprefix }}%
\providecommand \urlprefix  [0]{URL }%
\providecommand \Eprint [0]{\href }%
\providecommand \doibase [0]{http://dx.doi.org/}%
\providecommand \selectlanguage [0]{\@gobble}%
\providecommand \bibinfo  [0]{\@secondoftwo}%
\providecommand \bibfield  [0]{\@secondoftwo}%
\providecommand \translation [1]{[#1]}%
\providecommand \BibitemOpen [0]{}%
\providecommand \bibitemStop [0]{}%
\providecommand \bibitemNoStop [0]{.\EOS\space}%
\providecommand \EOS [0]{\spacefactor3000\relax}%
\providecommand \BibitemShut  [1]{\csname bibitem#1\endcsname}%
\let\auto@bib@innerbib\@empty
%</preamble>
\bibitem [{\citenamefont {Novoselov}\ \emph {et~al.}(2004)\citenamefont
  {Novoselov}, \citenamefont {Geim}, \citenamefont {Morozov}, \citenamefont
  {Jiang}, \citenamefont {Zhang}, \citenamefont {Dubonos}, \citenamefont
  {Grigorieva},\ and\ \citenamefont {Firsov}}]{KNovoselov:Science04}%
  \BibitemOpen
  \bibfield  {author} {\bibinfo {author} {\bibfnamefont {K.}~\bibnamefont
  {Novoselov}}, \bibinfo {author} {\bibfnamefont {A.}~\bibnamefont {Geim}},
  \bibinfo {author} {\bibfnamefont {S.}~\bibnamefont {Morozov}}, \bibinfo
  {author} {\bibfnamefont {D.}~\bibnamefont {Jiang}}, \bibinfo {author}
  {\bibfnamefont {Y.}~\bibnamefont {Zhang}}, \bibinfo {author} {\bibfnamefont
  {S.}~\bibnamefont {Dubonos}}, \bibinfo {author} {\bibfnamefont
  {I.}~\bibnamefont {Grigorieva}}, \ and\ \bibinfo {author} {\bibfnamefont
  {A.}~\bibnamefont {Firsov}},\ }\href@noop {} {\bibfield  {journal} {\bibinfo
  {journal} {Science}\ }\textbf {\bibinfo {volume} {306}},\ \bibinfo {pages}
  {666} (\bibinfo {year} {2004})}\BibitemShut {NoStop}%
\bibitem [{\citenamefont {Wang}\ \emph {et~al.}(2012)\citenamefont {Wang},
  \citenamefont {Kalantar-Zadeh}, \citenamefont {Kis}, \citenamefont
  {Coleman},\ and\ \citenamefont {Strano}}]{QHWang:NatureNano12}%
  \BibitemOpen
  \bibfield  {author} {\bibinfo {author} {\bibfnamefont {Q.}~\bibnamefont
  {Wang}}, \bibinfo {author} {\bibfnamefont {K.}~\bibnamefont
  {Kalantar-Zadeh}}, \bibinfo {author} {\bibfnamefont {A.}~\bibnamefont {Kis}},
  \bibinfo {author} {\bibfnamefont {J.}~\bibnamefont {Coleman}}, \ and\
  \bibinfo {author} {\bibfnamefont {M.}~\bibnamefont {Strano}},\ }\href@noop {}
  {\bibfield  {journal} {\bibinfo  {journal} {Nature Nanotechnology}\ }\textbf
  {\bibinfo {volume} {7}},\ \bibinfo {pages} {699} (\bibinfo {year}
  {2012})}\BibitemShut {NoStop}%
\bibitem [{\citenamefont {Radisavljevic}\ \emph {et~al.}(2011)\citenamefont
  {Radisavljevic}, \citenamefont {Radenovic}, \citenamefont {Brivio},
  \citenamefont {Giacometti},\ and\ \citenamefont
  {Kis}}]{BRadisavljevic:Nature11}%
  \BibitemOpen
  \bibfield  {author} {\bibinfo {author} {\bibfnamefont {B.}~\bibnamefont
  {Radisavljevic}}, \bibinfo {author} {\bibfnamefont {A.}~\bibnamefont
  {Radenovic}}, \bibinfo {author} {\bibfnamefont {J.}~\bibnamefont {Brivio}},
  \bibinfo {author} {\bibfnamefont {V.}~\bibnamefont {Giacometti}}, \ and\
  \bibinfo {author} {\bibfnamefont {A.}~\bibnamefont {Kis}},\ }\href@noop {}
  {\bibfield  {journal} {\bibinfo  {journal} {Nature Nanotechnology}\ }\textbf
  {\bibinfo {volume} {6}},\ \bibinfo {pages} {147} (\bibinfo {year}
  {2011})}\BibitemShut {NoStop}%
\bibitem [{\citenamefont {Yoon}\ \emph {et~al.}(2011)\citenamefont {Yoon},
  \citenamefont {Ganapathi},\ and\ \citenamefont {Salahuddin}}]{YYoon:NL11}%
  \BibitemOpen
  \bibfield  {author} {\bibinfo {author} {\bibfnamefont {Y.}~\bibnamefont
  {Yoon}}, \bibinfo {author} {\bibfnamefont {K.}~\bibnamefont {Ganapathi}}, \
  and\ \bibinfo {author} {\bibfnamefont {S.}~\bibnamefont {Salahuddin}},\
  }\href@noop {} {\bibfield  {journal} {\bibinfo  {journal} {Nano Lett.}\
  }\textbf {\bibinfo {volume} {11}},\ \bibinfo {pages} {3768} (\bibinfo {year}
  {2011})}\BibitemShut {NoStop}%
\bibitem [{\citenamefont {Kaasbjerg}\ \emph {et~al.}(2012)\citenamefont
  {Kaasbjerg}, \citenamefont {Thygesen},\ and\ \citenamefont
  {Jacobsen}}]{KKaasbjerg:PRB12}%
  \BibitemOpen
  \bibfield  {author} {\bibinfo {author} {\bibfnamefont {K.}~\bibnamefont
  {Kaasbjerg}}, \bibinfo {author} {\bibfnamefont {K.~S.}\ \bibnamefont
  {Thygesen}}, \ and\ \bibinfo {author} {\bibfnamefont {K.~W.}\ \bibnamefont
  {Jacobsen}},\ }\href@noop {} {\bibfield  {journal} {\bibinfo  {journal}
  {Phys. Rev. B}\ }\textbf {\bibinfo {volume} {85}},\ \bibinfo {pages} {115317}
  (\bibinfo {year} {2012})}\BibitemShut {NoStop}%
\bibitem [{\citenamefont {Lin}\ \emph {et~al.}(2012)\citenamefont {Lin},
  \citenamefont {Liu}, \citenamefont {Lan}, \citenamefont {Tan}, \citenamefont
  {Dhindsa}, \citenamefont {Zeng}, \citenamefont {Naik}, \citenamefont
  {Cheng},\ and\ \citenamefont {Zhou}}]{MWLin:JPD12}%
  \BibitemOpen
  \bibfield  {author} {\bibinfo {author} {\bibfnamefont {M.-W.}\ \bibnamefont
  {Lin}}, \bibinfo {author} {\bibfnamefont {L.}~\bibnamefont {Liu}}, \bibinfo
  {author} {\bibfnamefont {Q.}~\bibnamefont {Lan}}, \bibinfo {author}
  {\bibfnamefont {X.}~\bibnamefont {Tan}}, \bibinfo {author} {\bibfnamefont
  {K.~S.}\ \bibnamefont {Dhindsa}}, \bibinfo {author} {\bibfnamefont
  {P.}~\bibnamefont {Zeng}}, \bibinfo {author} {\bibfnamefont {V.~M.}\
  \bibnamefont {Naik}}, \bibinfo {author} {\bibfnamefont {M.~M.-C.}\
  \bibnamefont {Cheng}}, \ and\ \bibinfo {author} {\bibfnamefont
  {Z.}~\bibnamefont {Zhou}},\ }\href@noop {} {\bibfield  {journal} {\bibinfo
  {journal} {J. Phys. D: Appl. Phys.}\ }\textbf {\bibinfo {volume} {45}},\
  \bibinfo {pages} {345102} (\bibinfo {year} {2012})}\BibitemShut {NoStop}%
\bibitem [{\citenamefont {Lin}\ \emph {et~al.}(2013)\citenamefont {Lin},
  \citenamefont {Zhong}, \citenamefont {Zhong}, \citenamefont {Li},
  \citenamefont {Zhang},\ and\ \citenamefont
  {Chen}}]{JLin:APL13_MoS2Overlayer}%
  \BibitemOpen
  \bibfield  {author} {\bibinfo {author} {\bibfnamefont {J.}~\bibnamefont
  {Lin}}, \bibinfo {author} {\bibfnamefont {J.}~\bibnamefont {Zhong}}, \bibinfo
  {author} {\bibfnamefont {S.}~\bibnamefont {Zhong}}, \bibinfo {author}
  {\bibfnamefont {H.}~\bibnamefont {Li}}, \bibinfo {author} {\bibfnamefont
  {H.}~\bibnamefont {Zhang}}, \ and\ \bibinfo {author} {\bibfnamefont
  {W.}~\bibnamefont {Chen}},\ }\href@noop {} {\bibfield  {journal} {\bibinfo
  {journal} {Appl. Phys. Lett.}\ }\textbf {\bibinfo {volume} {103}},\ \bibinfo
  {pages} {063109} (\bibinfo {year} {2013})}\BibitemShut {NoStop}%
\bibitem [{\citenamefont {Radisavljevic}\ and\ \citenamefont
  {Kis}(2013)}]{BRadisavljevic:NatMat13}%
  \BibitemOpen
  \bibfield  {author} {\bibinfo {author} {\bibfnamefont {B.}~\bibnamefont
  {Radisavljevic}}\ and\ \bibinfo {author} {\bibfnamefont {A.}~\bibnamefont
  {Kis}},\ }\href@noop {} {\bibfield  {journal} {\bibinfo  {journal} {Nat.
  Mater.}\ }\textbf {\bibinfo {volume} {12}},\ \bibinfo {pages} {815} (\bibinfo
  {year} {2013})}\BibitemShut {NoStop}%
\bibitem [{\citenamefont {Adam}\ \emph {et~al.}(2009)\citenamefont {Adam},
  \citenamefont {Hwang}, \citenamefont {Rossi},\ and\ \citenamefont
  {Das~Sarma}}]{SAdam:SSC09}%
  \BibitemOpen
  \bibfield  {author} {\bibinfo {author} {\bibfnamefont {S.}~\bibnamefont
  {Adam}}, \bibinfo {author} {\bibfnamefont {E.}~\bibnamefont {Hwang}},
  \bibinfo {author} {\bibfnamefont {E.}~\bibnamefont {Rossi}}, \ and\ \bibinfo
  {author} {\bibfnamefont {S.}~\bibnamefont {Das~Sarma}},\ }\href@noop {}
  {\bibfield  {journal} {\bibinfo  {journal} {Solid State Commun.}\ }\textbf
  {\bibinfo {volume} {149}},\ \bibinfo {pages} {1072} (\bibinfo {year}
  {2009})}\BibitemShut {NoStop}%
\bibitem [{\citenamefont {Fischetti}\ \emph {et~al.}(2001)\citenamefont
  {Fischetti}, \citenamefont {Neumayer},\ and\ \citenamefont
  {Cartier}}]{MVFischetti:JAP01}%
  \BibitemOpen
  \bibfield  {author} {\bibinfo {author} {\bibfnamefont {M.~V.}\ \bibnamefont
  {Fischetti}}, \bibinfo {author} {\bibfnamefont {D.~A.}\ \bibnamefont
  {Neumayer}}, \ and\ \bibinfo {author} {\bibfnamefont {E.~A.}\ \bibnamefont
  {Cartier}},\ }\href@noop {} {\bibfield  {journal} {\bibinfo  {journal} {J.
  Appl. Phys.}\ }\textbf {\bibinfo {volume} {90}},\ \bibinfo {pages} {4587}
  (\bibinfo {year} {2001})}\BibitemShut {NoStop}%
\bibitem [{\citenamefont {Fratini}\ and\ \citenamefont
  {Guinea}(2008)}]{SFratini:PRB08}%
  \BibitemOpen
  \bibfield  {author} {\bibinfo {author} {\bibfnamefont {S.}~\bibnamefont
  {Fratini}}\ and\ \bibinfo {author} {\bibfnamefont {F.}~\bibnamefont
  {Guinea}},\ }\href@noop {} {\bibfield  {journal} {\bibinfo  {journal} {Phys.
  Rev. B}\ }\textbf {\bibinfo {volume} {77}},\ \bibinfo {pages} {195415}
  (\bibinfo {year} {2008})}\BibitemShut {NoStop}%
\bibitem [{\citenamefont {Ong}\ and\ \citenamefont
  {Fischetti}(2012{\natexlab{a}})}]{ZYOng:PRB12_IPP}%
  \BibitemOpen
  \bibfield  {author} {\bibinfo {author} {\bibfnamefont {Z.-Y.}\ \bibnamefont
  {Ong}}\ and\ \bibinfo {author} {\bibfnamefont {M.~V.}\ \bibnamefont
  {Fischetti}},\ }\href@noop {} {\bibfield  {journal} {\bibinfo  {journal}
  {Phys. Rev. B}\ }\textbf {\bibinfo {volume} {86}},\ \bibinfo {pages} {165422}
  (\bibinfo {year} {2012}{\natexlab{a}})}\BibitemShut {NoStop}%
\bibitem [{\citenamefont {Ong}\ and\ \citenamefont
  {Fischetti}(2012{\natexlab{b}})}]{ZYOng:PRB12_Erratum}%
  \BibitemOpen
  \bibfield  {author} {\bibinfo {author} {\bibfnamefont {Z.-Y.}\ \bibnamefont
  {Ong}}\ and\ \bibinfo {author} {\bibfnamefont {M.~V.}\ \bibnamefont
  {Fischetti}},\ }\href@noop {} {\bibfield  {journal} {\bibinfo  {journal}
  {Phys. Rev. B}\ }\textbf {\bibinfo {volume} {86}},\ \bibinfo {pages}
  {199904(E)} (\bibinfo {year} {2012}{\natexlab{b}})}\BibitemShut {NoStop}%
\bibitem [{\citenamefont {Ong}\ and\ \citenamefont
  {Fischetti}(2013{\natexlab{a}})}]{ZYOng:APL13_TopOxide}%
  \BibitemOpen
  \bibfield  {author} {\bibinfo {author} {\bibfnamefont {Z.-Y.}\ \bibnamefont
  {Ong}}\ and\ \bibinfo {author} {\bibfnamefont {M.~V.}\ \bibnamefont
  {Fischetti}},\ }\href@noop {} {\bibfield  {journal} {\bibinfo  {journal}
  {Appl. Phys. Lett.}\ }\textbf {\bibinfo {volume} {102}},\ \bibinfo {pages}
  {183506} (\bibinfo {year} {2013}{\natexlab{a}})}\BibitemShut {NoStop}%
\bibitem [{\citenamefont {Amani}\ \emph {et~al.}(2013)\citenamefont {Amani},
  \citenamefont {Chin}, \citenamefont {Birdwell}, \citenamefont {O'Regan},
  \citenamefont {Najmaei}, \citenamefont {Liu}, \citenamefont {Ajayan},
  \citenamefont {Lou},\ and\ \citenamefont {Dubey}}]{MAmani:APL13}%
  \BibitemOpen
  \bibfield  {author} {\bibinfo {author} {\bibfnamefont {M.}~\bibnamefont
  {Amani}}, \bibinfo {author} {\bibfnamefont {M.~L.}\ \bibnamefont {Chin}},
  \bibinfo {author} {\bibfnamefont {A.~G.}\ \bibnamefont {Birdwell}}, \bibinfo
  {author} {\bibfnamefont {T.~P.}\ \bibnamefont {O'Regan}}, \bibinfo {author}
  {\bibfnamefont {S.}~\bibnamefont {Najmaei}}, \bibinfo {author} {\bibfnamefont
  {Z.}~\bibnamefont {Liu}}, \bibinfo {author} {\bibfnamefont {P.~M.}\
  \bibnamefont {Ajayan}}, \bibinfo {author} {\bibfnamefont {J.}~\bibnamefont
  {Lou}}, \ and\ \bibinfo {author} {\bibfnamefont {M.}~\bibnamefont {Dubey}},\
  }\href@noop {} {\bibfield  {journal} {\bibinfo  {journal} {Appl. Phys.
  Lett.}\ }\textbf {\bibinfo {volume} {102}},\ \bibinfo {pages} {193107}
  (\bibinfo {year} {2013})}\BibitemShut {NoStop}%
\bibitem [{\citenamefont {Fallahazad}\ \emph {et~al.}(2010)\citenamefont
  {Fallahazad}, \citenamefont {Kim}, \citenamefont {Colombo},\ and\
  \citenamefont {Tutuc}}]{BFallahazad:APL10}%
  \BibitemOpen
  \bibfield  {author} {\bibinfo {author} {\bibfnamefont {B.}~\bibnamefont
  {Fallahazad}}, \bibinfo {author} {\bibfnamefont {S.}~\bibnamefont {Kim}},
  \bibinfo {author} {\bibfnamefont {L.}~\bibnamefont {Colombo}}, \ and\
  \bibinfo {author} {\bibfnamefont {E.}~\bibnamefont {Tutuc}},\ }\href@noop {}
  {\bibfield  {journal} {\bibinfo  {journal} {Appl. Phys. Lett.}\ }\textbf
  {\bibinfo {volume} {97}},\ \bibinfo {pages} {123105} (\bibinfo {year}
  {2010})}\BibitemShut {NoStop}%
\bibitem [{\citenamefont {Ong}\ and\ \citenamefont
  {Fischetti}(2012{\natexlab{c}})}]{ZYOng:PRB12_TopGate}%
  \BibitemOpen
  \bibfield  {author} {\bibinfo {author} {\bibfnamefont {Z.-Y.}\ \bibnamefont
  {Ong}}\ and\ \bibinfo {author} {\bibfnamefont {M.~V.}\ \bibnamefont
  {Fischetti}},\ }\href@noop {} {\bibfield  {journal} {\bibinfo  {journal}
  {Phys. Rev. B}\ }\textbf {\bibinfo {volume} {86}},\ \bibinfo {pages} {121409}
  (\bibinfo {year} {2012}{\natexlab{c}})}\BibitemShut {NoStop}%
\bibitem [{\citenamefont {Lemme}\ \emph {et~al.}(2008)\citenamefont {Lemme},
  \citenamefont {Echtermeyer}, \citenamefont {Baus}, \citenamefont {Szafranek},
  \citenamefont {Bolten}, \citenamefont {Schmidt}, \citenamefont {Wahlbrink},\
  and\ \citenamefont {Kurz}}]{MCLemme:SSE08}%
  \BibitemOpen
  \bibfield  {author} {\bibinfo {author} {\bibfnamefont {M.}~\bibnamefont
  {Lemme}}, \bibinfo {author} {\bibfnamefont {T.}~\bibnamefont {Echtermeyer}},
  \bibinfo {author} {\bibfnamefont {M.}~\bibnamefont {Baus}}, \bibinfo {author}
  {\bibfnamefont {B.}~\bibnamefont {Szafranek}}, \bibinfo {author}
  {\bibfnamefont {J.}~\bibnamefont {Bolten}}, \bibinfo {author} {\bibfnamefont
  {M.}~\bibnamefont {Schmidt}}, \bibinfo {author} {\bibfnamefont
  {T.}~\bibnamefont {Wahlbrink}}, \ and\ \bibinfo {author} {\bibfnamefont
  {H.}~\bibnamefont {Kurz}},\ }\href@noop {} {\bibfield  {journal} {\bibinfo
  {journal} {Solid-State Electron.}\ }\textbf {\bibinfo {volume} {52}},\
  \bibinfo {pages} {514} (\bibinfo {year} {2008})}\BibitemShut {NoStop}%
\bibitem [{\citenamefont {Kim}\ \emph {et~al.}(2009)\citenamefont {Kim},
  \citenamefont {Nah}, \citenamefont {Jo}, \citenamefont {Shahrjerdi},
  \citenamefont {Colombo}, \citenamefont {Yao}, \citenamefont {Tutuc},\ and\
  \citenamefont {Banerjee}}]{SKim:APL09}%
  \BibitemOpen
  \bibfield  {author} {\bibinfo {author} {\bibfnamefont {S.}~\bibnamefont
  {Kim}}, \bibinfo {author} {\bibfnamefont {J.}~\bibnamefont {Nah}}, \bibinfo
  {author} {\bibfnamefont {I.}~\bibnamefont {Jo}}, \bibinfo {author}
  {\bibfnamefont {D.}~\bibnamefont {Shahrjerdi}}, \bibinfo {author}
  {\bibfnamefont {L.}~\bibnamefont {Colombo}}, \bibinfo {author} {\bibfnamefont
  {Z.}~\bibnamefont {Yao}}, \bibinfo {author} {\bibfnamefont {E.}~\bibnamefont
  {Tutuc}}, \ and\ \bibinfo {author} {\bibfnamefont {S.}~\bibnamefont
  {Banerjee}},\ }\href@noop {} {\bibfield  {journal} {\bibinfo  {journal}
  {Appl. Phys. Lett}\ }\textbf {\bibinfo {volume} {94}},\ \bibinfo {pages}
  {062107} (\bibinfo {year} {2009})}\BibitemShut {NoStop}%
\bibitem [{\citenamefont {Pezoldt}\ \emph {et~al.}(2010)\citenamefont
  {Pezoldt}, \citenamefont {Hummel}, \citenamefont {Hanisch}, \citenamefont
  {Hotovy}, \citenamefont {Kadlecikova},\ and\ \citenamefont
  {Schwierz}}]{JPezoldt:PSS10}%
  \BibitemOpen
  \bibfield  {author} {\bibinfo {author} {\bibfnamefont {J.}~\bibnamefont
  {Pezoldt}}, \bibinfo {author} {\bibfnamefont {C.}~\bibnamefont {Hummel}},
  \bibinfo {author} {\bibfnamefont {A.}~\bibnamefont {Hanisch}}, \bibinfo
  {author} {\bibfnamefont {I.}~\bibnamefont {Hotovy}}, \bibinfo {author}
  {\bibfnamefont {M.}~\bibnamefont {Kadlecikova}}, \ and\ \bibinfo {author}
  {\bibfnamefont {F.}~\bibnamefont {Schwierz}},\ }\href@noop {} {\bibfield
  {journal} {\bibinfo  {journal} {Phys. Status Solidi C}\ }\textbf {\bibinfo
  {volume} {7}},\ \bibinfo {pages} {390} (\bibinfo {year} {2010})}\BibitemShut
  {NoStop}%
\bibitem [{\citenamefont {Zou}\ \emph {et~al.}(2010)\citenamefont {Zou},
  \citenamefont {Hong}, \citenamefont {Keefer},\ and\ \citenamefont
  {Zhu}}]{KZou:PRL10}%
  \BibitemOpen
  \bibfield  {author} {\bibinfo {author} {\bibfnamefont {K.}~\bibnamefont
  {Zou}}, \bibinfo {author} {\bibfnamefont {X.}~\bibnamefont {Hong}}, \bibinfo
  {author} {\bibfnamefont {D.}~\bibnamefont {Keefer}}, \ and\ \bibinfo {author}
  {\bibfnamefont {J.}~\bibnamefont {Zhu}},\ }\href@noop {} {\bibfield
  {journal} {\bibinfo  {journal} {Phys. Rev. Lett.}\ }\textbf {\bibinfo
  {volume} {105}},\ \bibinfo {pages} {126601} (\bibinfo {year}
  {2010})}\BibitemShut {NoStop}%
\bibitem [{\citenamefont {Xia}\ \emph {et~al.}(2010)\citenamefont {Xia},
  \citenamefont {Chen}, \citenamefont {Wiktor}, \citenamefont {Ferry},\ and\
  \citenamefont {Tao}}]{JLXia:NL10}%
  \BibitemOpen
  \bibfield  {author} {\bibinfo {author} {\bibfnamefont {J.~L.}\ \bibnamefont
  {Xia}}, \bibinfo {author} {\bibfnamefont {F.}~\bibnamefont {Chen}}, \bibinfo
  {author} {\bibfnamefont {P.}~\bibnamefont {Wiktor}}, \bibinfo {author}
  {\bibfnamefont {D.~K.}\ \bibnamefont {Ferry}}, \ and\ \bibinfo {author}
  {\bibfnamefont {N.~J.}\ \bibnamefont {Tao}},\ }\href@noop {} {\bibfield
  {journal} {\bibinfo  {journal} {Nano Lett.}\ }\textbf {\bibinfo {volume}
  {10}},\ \bibinfo {pages} {5060} (\bibinfo {year} {2010})}\BibitemShut
  {NoStop}%
\bibitem [{\citenamefont {Fuhrer}\ and\ \citenamefont
  {Hone}(2013)}]{MFuhrer:NatureNanotech13}%
  \BibitemOpen
  \bibfield  {author} {\bibinfo {author} {\bibfnamefont {M.~S.}\ \bibnamefont
  {Fuhrer}}\ and\ \bibinfo {author} {\bibfnamefont {J.}~\bibnamefont {Hone}},\
  }\href@noop {} {\bibfield  {journal} {\bibinfo  {journal} {Nature
  Nanotechnology}\ }\textbf {\bibinfo {volume} {8}},\ \bibinfo {pages} {146}
  (\bibinfo {year} {2013})}\BibitemShut {NoStop}%
\bibitem [{\citenamefont {Baugher}\ \emph {et~al.}(2013)\citenamefont
  {Baugher}, \citenamefont {Churchill}, \citenamefont {Yang},\ and\
  \citenamefont {Jarillo-Herrero}}]{BBaugher:NL13}%
  \BibitemOpen
  \bibfield  {author} {\bibinfo {author} {\bibfnamefont {B.}~\bibnamefont
  {Baugher}}, \bibinfo {author} {\bibfnamefont {H.~O.~H.}\ \bibnamefont
  {Churchill}}, \bibinfo {author} {\bibfnamefont {Y.}~\bibnamefont {Yang}}, \
  and\ \bibinfo {author} {\bibfnamefont {P.}~\bibnamefont {Jarillo-Herrero}},\
  }\href@noop {} {\bibfield  {journal} {\bibinfo  {journal} {Nano Lett.}\
  }\textbf {\bibinfo {volume} {13}},\ \bibinfo {pages} {4212} (\bibinfo {year}
  {2013})}\BibitemShut {NoStop}%
\bibitem [{\citenamefont {Ghatak}\ \emph {et~al.}(2011)\citenamefont {Ghatak},
  \citenamefont {Pal},\ and\ \citenamefont {Ghosh}}]{SGhatak:ACSNano11}%
  \BibitemOpen
  \bibfield  {author} {\bibinfo {author} {\bibfnamefont {S.}~\bibnamefont
  {Ghatak}}, \bibinfo {author} {\bibfnamefont {A.~N.}\ \bibnamefont {Pal}}, \
  and\ \bibinfo {author} {\bibfnamefont {A.}~\bibnamefont {Ghosh}},\
  }\href@noop {} {\bibfield  {journal} {\bibinfo  {journal} {ACS Nano}\
  }\textbf {\bibinfo {volume} {5}},\ \bibinfo {pages} {7707} (\bibinfo {year}
  {2011})}\BibitemShut {NoStop}%
\bibitem [{\citenamefont {Davies}(1997)}]{JHDavies:Book97}%
  \BibitemOpen
  \bibfield  {author} {\bibinfo {author} {\bibfnamefont {J.~H.}\ \bibnamefont
  {Davies}},\ }\href@noop {} {\emph {\bibinfo {title} {The Physics of
  Low-Dimensional Semiconductors}}}\ (\bibinfo  {publisher} {Cambridge
  University Press},\ \bibinfo {address} {Cambridge, UK},\ \bibinfo {year}
  {1997})\BibitemShut {NoStop}%
\bibitem [{\citenamefont {Maldague}(1978)}]{PMaldague:SurfSci1978}%
  \BibitemOpen
  \bibfield  {author} {\bibinfo {author} {\bibfnamefont {P.~F.}\ \bibnamefont
  {Maldague}},\ }\href@noop {} {\bibfield  {journal} {\bibinfo  {journal}
  {Surf. Sci.}\ }\textbf {\bibinfo {volume} {73}},\ \bibinfo {pages} {296}
  (\bibinfo {year} {1978})}\BibitemShut {NoStop}%
\bibitem [{\citenamefont {Stern}(1980)}]{FStern:PRL80}%
  \BibitemOpen
  \bibfield  {author} {\bibinfo {author} {\bibfnamefont {F.}~\bibnamefont
  {Stern}},\ }\href@noop {} {\bibfield  {journal} {\bibinfo  {journal} {Phys.
  Rev. Lett.}\ }\textbf {\bibinfo {volume} {44}},\ \bibinfo {pages} {1469}
  (\bibinfo {year} {1980})}\BibitemShut {NoStop}%
\bibitem [{\citenamefont {Ando}\ \emph {et~al.}(1982)\citenamefont {Ando},
  \citenamefont {Fowler},\ and\ \citenamefont {Stern}}]{TAndo:RMP82}%
  \BibitemOpen
  \bibfield  {author} {\bibinfo {author} {\bibfnamefont {T.}~\bibnamefont
  {Ando}}, \bibinfo {author} {\bibfnamefont {A.~B.}\ \bibnamefont {Fowler}}, \
  and\ \bibinfo {author} {\bibfnamefont {F.}~\bibnamefont {Stern}},\
  }\href@noop {} {\bibfield  {journal} {\bibinfo  {journal} {Rev. Mod. Phys.}\
  }\textbf {\bibinfo {volume} {54}},\ \bibinfo {pages} {437} (\bibinfo {year}
  {1982})}\BibitemShut {NoStop}%
\bibitem [{\citenamefont {Li}\ \emph {et~al.}(2013)\citenamefont {Li},
  \citenamefont {Mullen}, \citenamefont {Jin}, \citenamefont {Borysenko},
  \citenamefont {Buongiorno~Nardelli},\ and\ \citenamefont
  {Kim}}]{XLi:PRB13_IntrinsicMoS2}%
  \BibitemOpen
  \bibfield  {author} {\bibinfo {author} {\bibfnamefont {X.}~\bibnamefont
  {Li}}, \bibinfo {author} {\bibfnamefont {J.~T.}\ \bibnamefont {Mullen}},
  \bibinfo {author} {\bibfnamefont {Z.}~\bibnamefont {Jin}}, \bibinfo {author}
  {\bibfnamefont {K.~M.}\ \bibnamefont {Borysenko}}, \bibinfo {author}
  {\bibfnamefont {M.}~\bibnamefont {Buongiorno~Nardelli}}, \ and\ \bibinfo
  {author} {\bibfnamefont {K.~W.}\ \bibnamefont {Kim}},\ }\href@noop {}
  {\bibfield  {journal} {\bibinfo  {journal} {Phys. Rev. B}\ }\textbf {\bibinfo
  {volume} {87}},\ \bibinfo {pages} {115418} (\bibinfo {year}
  {2013})}\BibitemShut {NoStop}%
\bibitem [{\citenamefont {Das~Sarma}(1986)}]{SDasSarma:PRB85}%
  \BibitemOpen
  \bibfield  {author} {\bibinfo {author} {\bibfnamefont {S.}~\bibnamefont
  {Das~Sarma}},\ }\href@noop {} {\bibfield  {journal} {\bibinfo  {journal}
  {Phys. Rev. B}\ }\textbf {\bibinfo {volume} {33}},\ \bibinfo {pages} {5401}
  (\bibinfo {year} {1986})}\BibitemShut {NoStop}%
\bibitem [{\citenamefont {Jariwala}\ \emph {et~al.}(2013)\citenamefont
  {Jariwala}, \citenamefont {Sangwan}, \citenamefont {Late}, \citenamefont
  {Johns}, \citenamefont {Dravid}, \citenamefont {Marks}, \citenamefont
  {Lauhon},\ and\ \citenamefont {Hersam}}]{DJariwala:APL13}%
  \BibitemOpen
  \bibfield  {author} {\bibinfo {author} {\bibfnamefont {D.}~\bibnamefont
  {Jariwala}}, \bibinfo {author} {\bibfnamefont {V.~K.}\ \bibnamefont
  {Sangwan}}, \bibinfo {author} {\bibfnamefont {D.~J.}\ \bibnamefont {Late}},
  \bibinfo {author} {\bibfnamefont {J.~E.}\ \bibnamefont {Johns}}, \bibinfo
  {author} {\bibfnamefont {V.~P.}\ \bibnamefont {Dravid}}, \bibinfo {author}
  {\bibfnamefont {T.~J.}\ \bibnamefont {Marks}}, \bibinfo {author}
  {\bibfnamefont {L.~J.}\ \bibnamefont {Lauhon}}, \ and\ \bibinfo {author}
  {\bibfnamefont {M.~C.}\ \bibnamefont {Hersam}},\ }\href@noop {} {\bibfield
  {journal} {\bibinfo  {journal} {Appl. Phys. Lett.}\ }\textbf {\bibinfo
  {volume} {102}},\ \bibinfo {pages} {173107} (\bibinfo {year}
  {2013})}\BibitemShut {NoStop}%
\bibitem [{\citenamefont {Robinson}\ \emph {et~al.}(2010)\citenamefont
  {Robinson}, \citenamefont {LaBella}, \citenamefont {Trumbull}, \citenamefont
  {Weng}, \citenamefont {Cavelero}, \citenamefont {Daniels}, \citenamefont
  {Hughes}, \citenamefont {Hollander}, \citenamefont {Fanton},\ and\
  \citenamefont {Snyder}}]{JRobinson:ACSNano10}%
  \BibitemOpen
  \bibfield  {author} {\bibinfo {author} {\bibfnamefont {J.~A.}\ \bibnamefont
  {Robinson}}, \bibinfo {author} {\bibfnamefont {M.}~\bibnamefont {LaBella}},
  \bibinfo {author} {\bibfnamefont {K.~A.}\ \bibnamefont {Trumbull}}, \bibinfo
  {author} {\bibfnamefont {X.}~\bibnamefont {Weng}}, \bibinfo {author}
  {\bibfnamefont {R.}~\bibnamefont {Cavelero}}, \bibinfo {author}
  {\bibfnamefont {T.}~\bibnamefont {Daniels}}, \bibinfo {author} {\bibfnamefont
  {Z.}~\bibnamefont {Hughes}}, \bibinfo {author} {\bibfnamefont
  {M.}~\bibnamefont {Hollander}}, \bibinfo {author} {\bibfnamefont
  {M.}~\bibnamefont {Fanton}}, \ and\ \bibinfo {author} {\bibfnamefont
  {D.}~\bibnamefont {Snyder}},\ }\href@noop {} {\bibfield  {journal} {\bibinfo
  {journal} {ACS Nano}\ }\textbf {\bibinfo {volume} {4}},\ \bibinfo {pages}
  {2667} (\bibinfo {year} {2010})}\BibitemShut {NoStop}%
\bibitem [{\citenamefont {Hollander}\ \emph {et~al.}(2011)\citenamefont
  {Hollander}, \citenamefont {LaBella}, \citenamefont {Hughes}, \citenamefont
  {Zhu}, \citenamefont {Trumbull}, \citenamefont {Cavalero}, \citenamefont
  {Snyder}, \citenamefont {Wang}, \citenamefont {Hwang}, \citenamefont
  {Datta},\ and\ \citenamefont {Robinson}}]{MHollander:NL11}%
  \BibitemOpen
  \bibfield  {author} {\bibinfo {author} {\bibfnamefont {M.~J.}\ \bibnamefont
  {Hollander}}, \bibinfo {author} {\bibfnamefont {M.}~\bibnamefont {LaBella}},
  \bibinfo {author} {\bibfnamefont {Z.~R.}\ \bibnamefont {Hughes}}, \bibinfo
  {author} {\bibfnamefont {M.}~\bibnamefont {Zhu}}, \bibinfo {author}
  {\bibfnamefont {K.~A.}\ \bibnamefont {Trumbull}}, \bibinfo {author}
  {\bibfnamefont {R.}~\bibnamefont {Cavalero}}, \bibinfo {author}
  {\bibfnamefont {D.~W.}\ \bibnamefont {Snyder}}, \bibinfo {author}
  {\bibfnamefont {X.}~\bibnamefont {Wang}}, \bibinfo {author} {\bibfnamefont
  {E.}~\bibnamefont {Hwang}}, \bibinfo {author} {\bibfnamefont
  {S.}~\bibnamefont {Datta}}, \ and\ \bibinfo {author} {\bibfnamefont {J.~A.}\
  \bibnamefont {Robinson}},\ }\href@noop {} {\bibfield  {journal} {\bibinfo
  {journal} {Nano Lett.}\ }\textbf {\bibinfo {volume} {11}},\ \bibinfo {pages}
  {3601} (\bibinfo {year} {2011})}\BibitemShut {NoStop}%
\bibitem [{\citenamefont {Zhou}\ \emph {et~al.}(2007)\citenamefont {Zhou},
  \citenamefont {Gweon}, \citenamefont {Fedorov}, \citenamefont {First},
  \citenamefont {De~Heer}, \citenamefont {Lee}, \citenamefont {Guinea},
  \citenamefont {Neto},\ and\ \citenamefont {Lanzara}}]{SYZhou:NatMat07}%
  \BibitemOpen
  \bibfield  {author} {\bibinfo {author} {\bibfnamefont {S.}~\bibnamefont
  {Zhou}}, \bibinfo {author} {\bibfnamefont {G.-H.}\ \bibnamefont {Gweon}},
  \bibinfo {author} {\bibfnamefont {A.}~\bibnamefont {Fedorov}}, \bibinfo
  {author} {\bibfnamefont {P.}~\bibnamefont {First}}, \bibinfo {author}
  {\bibfnamefont {W.}~\bibnamefont {De~Heer}}, \bibinfo {author} {\bibfnamefont
  {D.-H.}\ \bibnamefont {Lee}}, \bibinfo {author} {\bibfnamefont
  {F.}~\bibnamefont {Guinea}}, \bibinfo {author} {\bibfnamefont {A.~C.}\
  \bibnamefont {Neto}}, \ and\ \bibinfo {author} {\bibfnamefont
  {A.}~\bibnamefont {Lanzara}},\ }\href@noop {} {\bibfield  {journal} {\bibinfo
   {journal} {Nat. Mater.}\ }\textbf {\bibinfo {volume} {6}},\ \bibinfo {pages}
  {770} (\bibinfo {year} {2007})}\BibitemShut {NoStop}%
\bibitem [{\citenamefont {Zhou}\ \emph {et~al.}(2008)\citenamefont {Zhou},
  \citenamefont {Siegel}, \citenamefont {Fedorov}, \citenamefont {El~Gabaly},
  \citenamefont {Schmid}, \citenamefont {Neto}, \citenamefont {Lee},\ and\
  \citenamefont {Lanzara}}]{SYZhou:NatMat08}%
  \BibitemOpen
  \bibfield  {author} {\bibinfo {author} {\bibfnamefont {S.}~\bibnamefont
  {Zhou}}, \bibinfo {author} {\bibfnamefont {D.}~\bibnamefont {Siegel}},
  \bibinfo {author} {\bibfnamefont {A.}~\bibnamefont {Fedorov}}, \bibinfo
  {author} {\bibfnamefont {F.}~\bibnamefont {El~Gabaly}}, \bibinfo {author}
  {\bibfnamefont {A.}~\bibnamefont {Schmid}}, \bibinfo {author} {\bibfnamefont
  {A.~C.}\ \bibnamefont {Neto}}, \bibinfo {author} {\bibfnamefont {D.-H.}\
  \bibnamefont {Lee}}, \ and\ \bibinfo {author} {\bibfnamefont
  {A.}~\bibnamefont {Lanzara}},\ }\href@noop {} {\bibfield  {journal} {\bibinfo
   {journal} {Nat. Mater.}\ }\textbf {\bibinfo {volume} {7}},\ \bibinfo {pages}
  {259} (\bibinfo {year} {2008})}\BibitemShut {NoStop}%
\bibitem [{\citenamefont {Ong}\ and\ \citenamefont
  {Fischetti}(2013{\natexlab{b}})}]{ZYOng:PRB13_RemotePhonon}%
  \BibitemOpen
  \bibfield  {author} {\bibinfo {author} {\bibfnamefont {Z.-Y.}\ \bibnamefont
  {Ong}}\ and\ \bibinfo {author} {\bibfnamefont {M.~V.}\ \bibnamefont
  {Fischetti}},\ }\href@noop {} {\bibfield  {journal} {\bibinfo  {journal}
  {Phys. Rev. B}\ }\textbf {\bibinfo {volume} {88}},\ \bibinfo {pages} {045405}
  (\bibinfo {year} {2013}{\natexlab{b}})}\BibitemShut {NoStop}%
\end{thebibliography}%

\end{document}